\documentclass[12pt]{iopart}
\pdfoutput=1

\usepackage{iopams}
\usepackage{bbm}
\usepackage{bm}
\usepackage[pdftex]{graphicx}
\usepackage{hyperref}

\expandafter\let\csname equation*\endcsname\relax

\expandafter\let\csname endequation*\endcsname\relax

\usepackage{amsmath}

\usepackage[caption=false]{subfig}
\usepackage{floatrow}
\usepackage{longtable}
\DeclareMathOperator*{\argmax}{arg\,max}

\newcommand{\strike}[1]{\ifmmode\setbox0\hbox{$#1$}%
\else
\setbox0\hbox{#1}%
\fi
\makebox[\the\wd0][c]{%
\rule[0.48\ht0]{0.5\wd0}{0.25pt}}\hspace*{-\the\wd0}#1}

\begin{document}

\title[Author guidelines for IOP Publishing journals in  \LaTeXe]{The cavity method for minority games between arbitrageurs on financial markets}

\author{Tim Ritmeester \& Hildegard Meyer-Ortmanns}

\address{Jacobs University Bremen, Campus Ring 1, 28759 Bremen, Germany}
\ead{t.ritmeester@jacobs-university.de; \qquad \;\;\;\;h.ortmanns@jacobs-university.de}
\vspace{10pt}
\begin{indented}
\item[]February 2022
\end{indented}

\begin{abstract}
We use the cavity method from statistical physics for analyzing the transient and stationary dynamics of a minority game that is played by agents performing market arbitrage. On the level of linear response the method allows to include the reaction of the market to individual actions of the agents as well as the reaction of the agents to individual information items of the market. This way we derive a self-consistent solution to the minority game. In particular we analyze the impact of general nonlinear price functions on the amount of arbitrage if noise from external fluctuations is present. We identify the conditions under which arbitrage gets reduced due to the presence of noise. When the cavity method is extended to time dependent response of the market price to previous actions of the agents, the individual contributions of noise can be pursued over different time scales in the transient dynamics until a stationary state is reached and when the stationary state is reached. The contributions are from external fluctuations in price and information and from noise due to the choice of strategies.  The   dynamics explains the time evolution of scores of the agents' strategies: it changes from initially a random walk to non-Markovian dynamics and bounded excursions on an intermediate time scale to effectively random switching in the choice between strategies on long time scales. In contrast to the Curie-Weiss level of a mean-field approach, the market response included by the cavity method captures the realistic feature that the agents can have a preference for a certain choice of strategies without getting stuck to a single choice. The breakdown of the method in the phase transition region indicates possible market mechanisms leading to critical volatility and a possible regime shift.
\end{abstract}

%
%
%
%
%

\section{Introduction}
 Conventional economic modeling of markets has relied on treating agents as rational decision makers,  homogenous to the extent that the impact of an individual agent on the aggregate system can be neglected \cite{bouchaud_economics_2008, bouchaud_unfortunate_2009}. From the physics perspective such models can be studied within the Curie-Weiss mean-field theory, for which  the macroscopic market is treated as a simple aggregate of the individual agents. However, empirically it is known that distributions of fluctuations in prices can have heavy tails, and such fluctuations may occur seemingly endogenously (i.e. without exogenous shocks) \cite{bouchaud_economics_2008, bouchaud_unfortunate_2009}. Models that are homogeneous with respect to the behaviour of the agents are unable to account for such effects.

In view of this,  the study of minority games \cite{challet_emergence_1997, challet2000modelling, coolen_minority_2005} provides an important development. Minority games are minimal models of financial markets when underlying minority mechanisms play a role. They describe the interaction between a large number of agents,
where each agent learns from the past to make strategic decisions in order to optimize its profits.
The agents interact through the market, such that interactions are effectively all-to-all. In common to minority games is that actions are only profitable if not too many other agents are involved in the same actions. This means that there is no rationally optimal decision. This requires the agents to diversify their strategies, leading to an inherently heterogenous system.

The minority game has led to a variety of novel insights,
including conditions under which prices behave as a random walk \cite{challet_phase_1999, chakraborti_statistical_2015} 
and general mechanisms which may lead the market to display scale-free fluctuations \cite{challet_criticality_2003, bouchaud_universal_2001}.
In \cite{ritmeester_minority_2021} we used the minority game to propose measures for controlling a dangerous abuse of arbitrage, as it most likely happened in 2019 in the energy market in Germany where it almost caused a nationwide blackout \cite{50_hertz_untersuchung_2019}. \\
Improved mean-field theories, developed in the context of spin glasses \cite{mezard_spin_1986}, provide insight into the dynamics of the minority game.
Under certain conditions, stationary states (that is, states in the long-time limit) of the minority game were shown to correspond  to the ground state of a spin glass Hamiltonian \cite{challet_phase_1999, marsili_continuum_2001}, similar to that of the Hopfield model.
In \cite{challet_statistical_2000} the replica method was used to solve for this ground state under a replica-symmetric ansatz, while the authors of \cite{coolen_dynamical_2001} used generating functionals to obtain a full dynamical solution of the minority game. These results explicitly showed that the minority game exhibits a replica-symmetry breaking transition. Such a transition leads to a phase containing long-range correlations, diverging relaxation times, and a proliferation of metastable states \cite{mezard_spin_1986, nishimori_statistical_2001, castellani_spin-glass_2005}.
From a physics point of view, it is of interest that mean-field theory of spin glasses may be utilized for minority games
with an intuitive interpretation of the presumed all-to-all interactions.

The cavity method provides an alternative to the replica method and to generating functionals for the analysis of spin glass models \cite{mezard_spin_1986, shamir_thouless-anderson-palmer_2000,mezard_mean-field_2017}. So far, the cavity method has not been used for analyzing the minority game despite the fact that the method has a natural interpretation in the context of financial or energy markets, as we shall see.
Compared to the Curie-Weiss mean-field theory, in which the impact of individual agents (here arbitrageurs) on prices is neglected, the cavity method implements an improvement in which the impact of an individual agent on the prices and vice versa, of an individual information item of the market on the agents, is treated in linear response. Quantification of the price impact of agents in financial markets is itself a well-studied topic \cite{bucci_crossover_2019,cont_price_2010}. By implementing linear response of the agents to the market and vice versa, we show that the cavity method leads to a self-consistent solution of the minority game, valid for the counterpart of the replica-symmetric phase. The solution quantifies the mutual impact of the agents on the market and leads to an explicit connection between the microscopic agents and the resulting macroscopic behaviour of the market. \\
In this work we use the cavity method both on the level of the stationary state, as indicated before, and on the level of the transient  dynamics towards the stationary state. In the stationary state we give the solution to the general non-linear minority game, previously analyzed only for anti-symmetric price functions \cite{papadopoulos_theory_2010}.
For general price functions, external and internal noise may interact both with each other and with the nonlinearities of the price function,  here we explicitly quantify this interaction beyond numerical results. At the level of the transient dynamics, we show that the impact of the agents on the prices combines with external and internal noise to give rise to three distinct time-scales in the time evolution of the choice of strategies. At short time-scales, external noise and noise due to random external information dominate, and the score of the strategies performs a random walk. At intermediate time-scales the scores follow a non-Markovian stochastic differential equation that we derive. At long time-scales, agents' switching between the two strategies is effectively described as binary noise: while external noise and noise due to random external fluctuations drive the scores of the strategies randomly, the market impact keeps the scores of different strategies close to each other, letting the agents switch between trading strategies.

The paper is organized as follows. In Sec.~\ref{sec: minority_game} we define the minority game for market arbitrage.
We review some results, and summarize the background information that is needed for applying the cavity method.
In Sec.~\ref{sec: linear_cavity} we solve the replica-symmetric phase of the non-linear minority game by the cavity method, while in Sec.~\ref{sec: dynamics}
we study the transient time evolution of the choice of best strategies, in particular the interaction between the agents' impact on prices and price fluctuations caused by both external noise and internal noise due to the provided market information. In the appendices we present some details of the calculations and a table of the notations and the most relevant variables used in this paper.

\section{The Minority Game for Market Arbitrage} \label{sec: minority_game}
\subsection{Definition of the minority game} \label{sec: minority}
In the context of financial markets, arbitrage is defined as the practice of taking advantage of price differences between two markets, by buying in one market and selling in the other (or vice-versa). In the minority game that we considered in \cite{ritmeester_minority_2021}, the agents were so-called balancing responsible parties (BRPs) who trade on behalf of retailers or producers on the energy market, but in the model act exclusively as arbitrageurs. In the present paper, the agents are again arbitrageurs who may act on general financial markets, not restricted to the energy market. When their actions are treated as a minority game, $N \gg 1$ agents make arbitrage actions $a^t_i = \pm 1$ (where $i$ denotes the agent) at times $t = 1, 2,\dots$. The agents attempt to maximize their pay-off, given for each $t$ by $- a_i^t \, g(A^t + \eta^t)$. The function $g$ is called the price function. For general financial markets, $g$ models the dependence of financial returns on the excess demand (i.e. on the difference between supply and demand), as discussed in \cite{coolen_minority_2005,bonnet_can_2010}. If the price of the asset on which arbitrage is performed at time $t$ is given by $\text{price}_t$, then:
\begin{align}
    \log(\text{price}_{t}) - \log(\text{price}_{t-1}) = g(A^t + \eta^t) \,.
\end{align}
Here $A^t + \eta^t$ is the excess demand, decomposed into a term due to arbitrage $A^t$ and a noise term $\eta^t$, which serves to model trading unrelated to arbitrage \cite{challet2000modelling, ritmeester_minority_2021}. In this sense we will refer to the noise $\eta^t$ as 'external'. In \cite{ritmeester_minority_2021} we made use of the fact  that specifically for the reserve energy market $g(A^t + \eta^t)$ corresponds directly to the $ \text{price}_{t}$, based on data from the German and Austrian energy markets we constructed concrete realizations of the function $g$. Here we will assume that $g' > 0$, $\eta^t$ is white noise drawn from a given distribution $P(\eta)$ (with zero mean). $A^t$ is the total amount of arbitrage, given by the rescaled sum of the agents' actions:
\begin{align}
    A^t = \frac{1}{\sqrt{N}}\sum_i a_i^t  \,.
\end{align}
The factor $\frac{1}{\sqrt{N}}$ keeps $A$ finite in the limit $N \rightarrow \infty$.
The fact that $g' > 0$ makes the game a minority game:
The more agents choose an action $a = +1$  ($-1$), the higher (lower) $g(A)$ becomes,
and the more profitable it becomes to choose the \textit{opposite} action.

\subsection{Limiting cases}
Let us first summarize from \cite{ritmeester_minority_2021} what kind of behaviour these pay-offs lead to.
If $g > 0$, the agents prefer the action $a = -1$, which would lead to $g$ being lowered again.
If $g < 0$ the oppositive happens. Hence, in the absence of any fluctuations, one would expect the agents to reach an equilibrium such that $g = 0$. This would imply that a fraction of $\frac{1}{2} \pm \frac{1}{2}b/\sqrt{N}$ of agents play $a = \pm 1$, where $b$ is a bias such that $g(b) = 0$. However, this requires the agents to achieve 'perfect anti-coordination': the agents should self-organize into two groups with exactly the right proportions. It is not clear without knowing the decisions about their competitors how the agents would achieve this. The opposite, more realistic scenario is one in which each agent acts completely independently, choosing randomly values from $a = \pm 1$ with probabilities $\frac{1}{2} \pm \frac{1}{2}b/\sqrt{N}$. This achieves on average $A = b$, but with some level of fluctuations around this average. Although the time average of $g(A^t)$ is expected to equal zero, generally this does not imply $g(b) = 0$ as the fluctuations of $A^t$ around $b$ may shift the average value of $g(A^t)$. This limiting case assumes that the agents act completely independently, and achieve no anti-coordination at all. In reality, over time the agents might learn to predict each others behaviour, and therefore reduce the level of fluctuations as compared to the case with no anti-coordination. The fluctuations in the amount of arbitrage $A^t$ and in the prices $g(A^t + \eta^t)$ are important to quantify, as they correspond to volatility in the market, which may have negative effects. In general financial markets they pose a risk to investors, while in the energy market they pose a danger for the stability of the power grid \cite{ritmeester_minority_2021}. 

\subsection{Agent-based modeling}
To give accurate predictions of the fluctuations as well as of the expected value of arbitrage $A^t$, we need to equip the agents with some explicit tactic to choose whether at any given time the best action is $a^t = 1$ or $-1$.
The agents wish to predict the actions of the other agents, before making their own decisions. In particular, this requires heterogeneity in the decision making between the agents; if all agents would have the same tactic,  making the same decisions,  hence all would obtain a bad pay-off. Several version of minority game learning dynamics have been studied; here we describe the most common version, which equips the agents with a basic form of reinforcement learning \cite{coolen_minority_2005, marsili_continuum_2001}. At each time $t$ a signal $\mu^t$ is given to the agents:
this is an abstract and compact representation of external circumstances such as the weather,
current events, market history, etc.
Here we assume $\mu^t$ is drawn uniformly at random (and independently for each $t$) from $P = O(N)$ possibilities $\mu^t = 1, \dots, P$.
The agents use these signals to predict which action $a = \pm 1$ is best.
They use strategies, each of which is a function that maps $\mu \rightarrow a = \pm 1$.
At the start of the game, each agent is assigned $S$ strategies out of all $2^P$ possible strategies. These are assigned by drawing each of the $P$ entries $s(\mu) = \pm 1$ randomly with probabilities $\frac{1}{2} \pm \frac{1}{2}b/\sqrt{N}$, for each $s$ from the set of $S$ strategies, where $b$ is a biasing parameter. $P$ and $S$ thus provide abstract parameterizations of the extent to which the arbitrageurs keep track of (external) information and of the extent to which they are able to (computationally) use this information, respectively. In Sec.~4 we will see that $P$ sets an effective time-scale of the minority game, as it determines the maximal diversity of information items $\mu$ that agents can explore. (Vice versa, analyzing time scales inherent in given time series of the  market, the complexity of the available information may be responsible for setting one of the time scales.) To decide which strategies to choose, the agents keep track of scores $U_s^t$ for each strategy $s$:
\begin{align}
    U_s^t &= \sum_{t' \leq t} -s(\mu^{t'})g_{t'} \,, \label{eq: score_definition} \\
    g_t & \equiv g(A^t + \eta^t)
\end{align}
which is the total pay-off that the strategy would have obtained \textit{in hindsight}.
The agents do not take into account their own impact on the market, since the actual choice $s'(\mu^{t^\prime})$ that agent $i$ made at time $t'$ and that enters $A^{t^\prime}$ in $g$ may differ from the evaluated $s(\mu^{t^\prime})$ if $s\not=s'$; a more sophisticated agent would take into account its own influence on $A^t$ and on the response of the other agents.
At each time step the agents then use the scores to decide on their best strategy $s_i^t$ and the corresponding action:
\begin{align}
    s_i^t &= \argmax_{s \in \bm{S}_i} U_s^{t-1}\,, \\
    a_i^t &= s_i^t(\mu^t) \,.
\end{align}
where $\bm{S}_i$ is the set of $S$ strategies available to $i$, this set is in general agent-dependent. The possibility of multiple strategies having equal score is avoided by assigning initial conditions $U_s^{t = 0}$ randomly, with $|U_s^{t = 0}| \ll 1$ (in practice we draw them from a normal distribution with standard deviation $10^{-10}$).

\subsection{Typical behaviour of the minority game} \label{sec: typical_behaviour}
\begin{figure}
    \centering
    \hspace{-0.45cm}
    \subfloat[]{\includegraphics[width = 0.45 \textwidth]{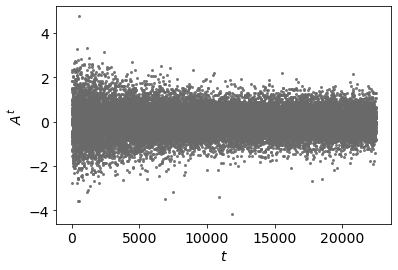}}
 \caption{The evolution of $A^t$, for $N = 4100$, $P = 2050$ and $S = 2$, showing the decrease in its fluctuations as a function of time.}
    \label{fig: scatteR_g}
\end{figure}

\begin{figure}
    \centering
    \hspace{-0.45cm}
    \begin{tabular}{ll}
    \subfloat[]{\includegraphics[width = 0.45 \textwidth]{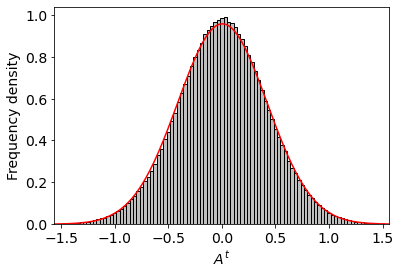}}
    &
    \subfloat[]{\includegraphics[width = 0.48 \textwidth]{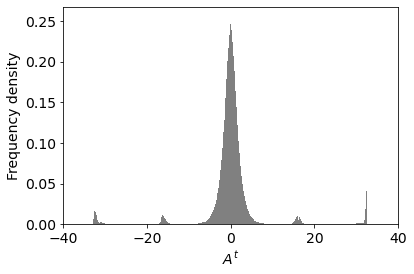}}
    \end{tabular}
    \caption{Histograms of $A^t$ (after convergence) for (a) $P = 2050$ and (b) $P = 32$; $N = 4100$, averages over $10^6$ time-steps. For low $P$ (b), the distribution is strongly non-Gaussian, in contrast to the high $P$ case (a). The qualitatively change is due to a phase transition at $P/N \approx 0.3374$.}
    \label{fig: histogram}
\end{figure}

\begin{figure}
    \centering
    \hspace{0.2cm}
    \begin{tabular}{ll}
    \subfloat[]{\includegraphics[width = 0.45 \textwidth]{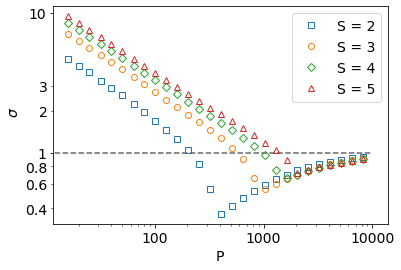}}
    &
    \subfloat[]{\includegraphics[width = 0.45 \textwidth]{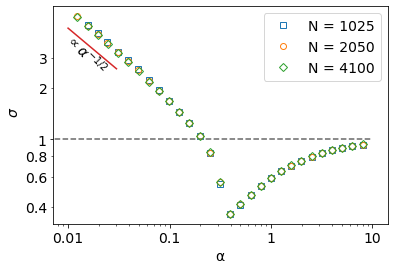}}
    \end{tabular}
    \caption{(a) Volatility for $N = 1025$, as a function of $P$ for different $S$. Each data-point is an average over $100$ samples. The dashed line corresponds to the hypothetical case, where agents would not have learned at all. (b) volatility for $S = 2$. The curves for different $N$ collapse on each other if they are shown as a function of $\alpha \equiv P/N$.}
    \label{fig: different_S}
\end{figure}

The type of behaviour resulting from the agent-based modeling is well-known \cite{challet2000modelling, coolen_minority_2005,ritmeester_minority_2021}, but it is instructive to review it.
As a starting point we consider particular values of the parameters: $N = 4100$, $P = 2050$, $S = 2$, zero noise and zero bias, and linear price function $g(A) = A$, as we considered in \cite{ritmeester_minority_2021}.
The time evolution of $A^t$ is shown in Fig.~\ref{fig: scatteR_g}.
The precise value of $A^t$ at a given time-step is highly unpredictable, but a general trend in the time evolution of $A$ (Fig.~\ref{fig: scatteR_g}) can be observed:
For small $t$, values of $A$ that are far from the average value are relatively common, indicating random decision making. As time evolves the agents learn to better anticipate $A$, and strong deviations  become more rare, until the distribution of $A^t$ becomes stationary.
Plotting a histogram of all the values attained by $A^t$, Fig.~\ref{fig: histogram}a shows that at large times $t$ the values of $A^t$ follow a Gaussian distribution.
The mean of the histogram is given by the bias $\overline{A^t + \eta^t} = \overline{A^t} = b$, while the width of the histogram depends on the parameters of the market, and is given by the volatility of the market defined as:
\begin{alignat}{1}
    \sigma \equiv \sqrt{\overline{(A^t  + \eta^t - b)^2}} \,.
\end{alignat}
Here the overline denotes a time average, defined as:
\begin{align}
    \overline{\mathbbm{O}_t} \equiv \lim_{t \rightarrow \infty} \frac{1}{t}\sum_{t' \leq t} \mathbbm{O}_{t'} \,.
\end{align}
The histogram shown in Fig.~\ref{fig: histogram}b shows the distribution of $A$ for a lower value of $P$ given by $P = 32$. This histogram is no longer Gaussian, and has rather extreme outliers. Lowering $P$ has thus given rise to qualitatively different behaviour. The reason is an underlying phase transition at a critical value of $\alpha=N/P$ as known from \cite{coolen_minority_2005}.
To investigate the dependence of the volatility on the parameters,
we run the minority game for different values of $N$, $P$ and $S$.
For any given set of parameters, we run the minority game until $\sigma$ converges to a constant value.
As the strategies are drawn randomly (corresponding to 'quenched disorder' in spin glasses), in principle even for the same parameters the volatility $\sigma$  varies at each time the game is played.
Therefore we repeat the procedure $100$ times;
the average values of $\sigma$ to which the system converges are shown in Fig.~\ref{fig: different_S}a for fixed $N$,
and varying values of $P$ and $S$.
For each of the values of $S$, the volatility $\sigma$ shows a non-trivial dependence on $P$.
For high $P$, the agents fail to learn at all,
and behave equivalently to deciding between actions at random ($\sigma = 1$).
As $P$ is lowered, the volatility also decreases,
until finally a turning point is reached, where $\sigma$ shoots up,
eventually reaching values much higher than the agents would have achieved if they had not learned at all.

Different values of $N$ are easily incorporated. As it turned out \cite{coolen_minority_2005}, $\sigma$ depends only on the ratio $\alpha \equiv P/N$;
this is shown for $S = 2$ in Fig.~\ref{fig: different_S}b.
The same holds for $S > 2$.
The non-monotonic behaviour is due to a phase transition at $\alpha_c \approx 0.3374$ (\cite{coolen_dynamical_2001}), corresponding to replica-symmetry breaking in spin glasses.
While for $\alpha > \alpha_c$ the histogram of $A^t$ is Gaussian, as in Fig.~\ref{fig: histogram}a, the low-$\alpha$ phase shows qualitatively different behaviour.

Now that we have summarized the resulting behaviour of the minority game,
we ask the following questions:
\begin{itemize}
    \item In what sense do the agents reach a stationary state at large times $t$, even if they keep on changing their strategies?
    \item Can we analytically derive the Gaussian distribution of the histogram of $A^t$ 
        and understand the nature of the phase transition?
    \item How do fluctuations in $A^t + \eta^t$ combine with a nonlinear price function $g$, to change the expected amount of arbitrage?
    \item What is the impact of an individual agent on the macroscopic behaviour of $A^t$?
    \item The agents update the scores of their strategies (Eq.~\ref{eq: score_definition}) in hindsight. Can we quantify the error they make in neglecting their own impact on the market?
\end{itemize}
The first two questions have been answered by the replica method and by generating functionals \cite{challet2000modelling,coolen_minority_2005,marsili_continuum_2001}.
We will reproduce these results  by the cavity method and use the cavity method to address also the last three questions. Together with the geometric interpretation of the phase transition given in \cite{challet2000modelling, chakraborti_statistical_2015} this leads to an in-depth understanding of the (high-$\alpha$ phase of) the minority game.

\subsection{Decomposition into three sources of volatility} \label{sec: sources_of_fluctuations}
Since we will pursue the transient competition of different sources of noise in Sec.~4, we first summarize the three independent sources and consider the case of $S=2$ for simplicity.
We denote the two strategies of each agent $i$ as  $s_i^\uparrow$ and $s_i^\downarrow$, respectively.
We then define the abbreviations
\begin{align}
\omega_i^\mu &\equiv \frac{s_i^\uparrow(\mu) + s_i^\downarrow(\mu)}{2} \,, \\
\xi_i^\mu &\equiv \frac{s_i^\uparrow(\mu) - s_i^\downarrow(\mu)}{2} \,, \\
U^t_i &\equiv U^t_{s_i^\uparrow} - U^t_{s_i^\downarrow} \,, \\
x^t_i &\equiv  \delta_{s_i^t, s_i^\uparrow} - \delta_{s_i^t, s_i^\downarrow} = \text{sign}(U^t_i) \,.
\end{align}
In terms of  these definitions, the action of an agent $i$ at time $t$ is given as $a_i^t = \omega_i^\mu + \xi_i^\mu x_i^t$.

The quantity $A^t + \eta^t$ fluctuates both because of the external noise $\eta$ and because of the fluctuating actions of the agents $a_i^t = \omega_i^\mu + \xi_i^\mu x_i^t$. These actions in turn fluctuate due to the random external information $\mu$ and due to the agents switching between the strategy that they use, i.e. between $x_i^t = \pm 1$. As in \cite{challet2000modelling, marsili_continuum_2001} we separate these latter two sources of fluctuations and define the quantities
\begin{align}
    A_\mu^t &\equiv \frac{1}{\sqrt{N}}\sum_i (\omega_i^{\mu}  + \xi_i^{\mu} x^t_i)\,,
\end{align}
where $\mu$ does not necessarily equal $\mu^t$. Since each $A_\mu^t$ is evaluated for fixed $\mu$, the only fluctuations in $A_\mu^t$ are due to fluctuations $x^t_i - x_i$, where $x_i$ is defined as $x_i = \overline{x_i^t}$.
In particular, the authors of \cite{challet2000modelling, marsili_continuum_2001} show that for a modification of the learning dynamics, in the high-$\alpha$ phase fluctuations in the strategy that an agent $i$ uses are uncorrelated to fluctuations in the strategy that any other agent $j$ uses: $\overline{(x^t_{i} - x_i) (x^t_{j} - x_j)} = \overline{(x^t_{i} - x_i)}\: \cdot \:\overline{(x^t_{j} - x_j)}$. We will re-derive this statement by means of the cavity method in Sec.~\ref{sec: dynamics} also for the unmodified learning dynamics, but use it here already to understand the fluctuations in $A_\mu^t$. We have:
\begin{align}
    A_\mu^t = \frac{1}{\sqrt{N}}\sum_i (\omega_i^\mu + \xi_i^\mu x_i^t) &= \frac{1}{\sqrt{N}}\sum_i (\omega_i^\mu + \xi_i^\mu x_i) + \frac{1}{\sqrt{N}}\sum_i \xi_i^\mu (x_i^t - x_i)  \\
    &\equiv A_\mu + \frac{1}{\sqrt{N}}\sum_i \xi_i^\mu (x_i^t - x_i)\,, \label{eq: A_mu_def}
\end{align}
where the last equation defines $A_\mu$. The first term is constant in time, while the second one fluctuates due to the agents switching between their strategies. Now, due to the initial random drawing of the strategies, the second term consists of random variables, which in the high-$\alpha$ phase are uncorrelated since $x_i^t - x_i$ and $x_j^t - x_j$ are uncorrelated. We can apply the central limit theorem to write \footnote{Denoting an average over the initial random drawing of the strategies as $\langle \dots \rangle_{s}$, what enters further in this result is the fact that the fluctuation $(x_i^t - x_i)$ is uncorrelated to $\xi_i^\mu$ and that $ \langle (\xi_i^\mu)^2 \rangle_s = \frac{1}{2} - \frac{b^2}{N} = \frac{1}{2} + O(1/N)$. Further, due to the central limit theorem the sums with and without self-averaging over the quenched disorder $\langle...\rangle_s$ differ by $O(1/\sqrt{N})$ and are therefore set equal.}:
%
\begin{align}
    (A_\mu^t - A_\mu) &\sim N(0, 1/2 - q_x/2) \,, \label{eq: A_mu_gaussian} \\
q_x &\equiv \frac{1}{N}\sum_i x_i^2 \,. \label{eq: q_x}
\end{align}
The volatility $\sigma$ (numerically determined in Fig.~\ref{fig: different_S}) then additively combines three sources of fluctuations: 
\begin{align}
    \sigma^2 \equiv \overline{(A^t + \eta^t  - b)^2} &= \sigma_\eta^2 + \frac{1}{P}\sum_\mu \overline{(A_\mu^t - b)^2} =  \sigma_\eta^2 + (1/2 - q_x/2) + q_A \,, \label{eq: sigma} \\
    q_A &\equiv \frac{1}{P} \sum_\mu (A_\mu - b)^2\,, \label{eq: q_A}
\end{align}
where $\sigma_\eta^2$ is the variance of the external noise, the contribution of $q_A$ describes the volatility due to random drawing of the signals $\mu$, and the term $(1/2 - q_x/2)$ describes the effective noise due to agents switching between their strategies. Quantification of the strength of the two sources of internal noise (resulting from the choice of strategies and from $\mu$) requires to calculate the associated quantities $q_x$ and $q_A$. The quantities $q_x$ and $q_A$ are macroscopic observables and play the role of order parameters in the related spin-glass theory. $q_x$ is a measure for the lack of diversity of strategies used by the agents (equal to $0$ if each agent has no preference between strategies, equal to $1$ if each agent uses only one of its strategies), while $q_A$ measures the extent to which the information $\mu$ predicts the market in the sense that as long $\mu$ causes fluctuations the agents want to exploit them for predictions ($q_A=0$ if $\mu$ does not provide a prediction for $A^t$, i.e. if all $A_\mu = b$). 

\subsection{Stationary state equations} \label{sec: stat_eq}
In Sec.~3 we want to derive the solutions of the stationary state equations in terms of the system parameters by means of the cavity method. This derivation is new and complementary to a derivation based on the replica method. The stationary state equations themselves have been derived in  \cite{marsili_continuum_2001}. Here we briefly re-derive these equations to fix  the notations. The equations refer to the time-averaged price function $g_\mu \equiv  \overline{g(A_\mu^t + \eta^t)}$ (for each given fixed $\mu$) and the variables $\{x_i\}$, characterizing each agent $i$ by the fraction of time in which  it uses each of its two strategies:
\begin{align}
    x_i = \overline{x_i^t} \in [-1,1]\,.
\end{align}
Stated differently, the variable $x_i \in [-1,1]$ describes the preference of agent $i$ between its two strategies $s_i^\uparrow$ and $s_i^\downarrow$:
if $x_i = 1$ ($x_i = -1$), the agent only uses its strategy $s^\uparrow_i$ ($s^\downarrow_i$), while for $x_i \approx 0$ the agent uses both strategies regularly and thus does not have a strong preference.
If an agent $i$ has $x_i = \pm 1$, this means that (in the long time limit) the agent uses only one of its two strategies. In this case we say the agent is 'frozen'. On the other hand, if a given agent $i$ regularly switches between strategies ($-1 < x_i < 1$), this requires the scores of the strategies to remain close to each other, i.e. $U_i^t$ to remain close to zero, such that $x_i^t = \text{sign}(U_i^t)$ can alternate. Formally:
\begin{align} \label{eq: stationary_state_a}
    \lim_{t \rightarrow \infty} U_i^t/t = 0  \qquad & \text{if} \qquad -1 < x_i < 1\\
    \lim_{t \rightarrow \infty}  U_i^t/t > 0 \qquad & \text{if}  \qquad x_i = 1 \\
    \lim_{t \rightarrow \infty}  U_i^t/t < 0 \qquad & \text{if}  \qquad x_i = -1 .
      \label{eq: stationary_state_b}
\end{align}
If the first condition does not hold, $U_i^t \equiv U^t_{s_i^\uparrow} - U^t_{s_i^\downarrow} $ will diverge in the limit $t \rightarrow \infty$, so the agent will never use the strategy with the lower score (leading to the second condition). The variables $g_\mu$ are introduced by writing:
\begin{equation}\label{eq: Ut/t}
    \lim_{t \rightarrow \infty} U_i^t/t = \overline{-\xi_i^{\mu^t}g_t}
    = \frac{1}{P} \sum_\mu - \xi_i^{\mu} g_{\mu}\, .
\end{equation}
Note that the time average over time dependent $\mu^t$ amounts to a sum over all $\mu$, each $\mu$ entering with weight $1/P$ due to the uniform distribution of $\mu$-values. In Sec.~\ref{sec: sources_of_fluctuations} we quantified the fluctuations in $A_\mu^t - A_\mu$. We then define $\delta = A_\mu^t - A_\mu$ to express $g_\mu$ in terms of $\{x_i\}$ as:
\begin{align}
    g_\mu \equiv  \overline{g(A_\mu^t + \eta^t)} &=  \int \mathrm{d}\delta \mathrm{d}\eta P(\delta)P(\eta) g(A_\mu + \delta + \eta), \\
    P(\delta) &\sim N\big(0, 1/2 - q_x/2\big) \, ,
\end{align}
where the dependence on $\{x_i\}$ enters through $A_\mu$ and $q_x$.
Note that here the time average over the price function is rewritten as an integral over two distributions of noise: The distribution $P(\delta) \sim N(0, 1/2 - q_x/2)$ describes the internal noise due to the agents' switching between strategies, while $P(\eta)$ is the distribution of external noise $\eta$. Let us write the average over these two types of noise as:
\begin{align}
    \big\langle \dots \big\rangle _{\delta, \eta} \equiv \int \mathrm{d}\delta \mathrm{d}\eta \dots P(\delta)P(\eta) \,.
\end{align}
In this notation, $g_{\mu} = \big\langle \, g(A_{\mu} + \delta + \eta) \,  \big\rangle_{\delta, \eta}$.

Thus, the stationary state equations for  $\{x_i\}$ and $\{g_\mu\}$ are given by:
\begin{alignat}{1}
    \forall i: \quad & -\frac{1}{\sqrt{N}}\sum_{\mu} \xi_i^\mu g_{\mu} =
    \begin{cases}
        0  &\text{if} \quad -1 < x_i < 1 \,, \\
        > 0 & \text{if} \quad x_i = 1 \,,
        \\
        < 0 & \text{if} \quad x_i = -1 \,.
    \end{cases} \label{eq: stat_eq_3}\\
    \forall \mu: \quad & \qquad \qquad \qquad \! \! g_{\mu} = \langle g(A_\mu + \delta + \eta)\rangle_{\delta, \eta} \,, \label{eq: stat_eq_g} \\
   & \qquad \qquad \qquad \! \!  A_{\mu} = \frac{1}{\sqrt{N}}\sum_i (\omega_i^\mu + \xi_i^\mu x_i)  \label{eq: stat_eq_A}\,,
\end{alignat}
where we rescaled both sides of Eq.~\ref{eq: stat_eq_3} by $\sqrt{\alpha P}$ for later convenience, leaving the solution unchanged. These stationary state equations were obtained first in \cite{marsili_continuum_2001}.
To complete the stationary state equations, we should fix the bias $b$.
In general we expect that the agents effectively adjust their bias such that we always have
\begin{equation}
\overline{g_t} = \frac{1}{P}\sum_{\mu} g_{\mu} = 0 \label{eq: stat_eq_b}
\end{equation}
(see \cite{challet_shedding_2004} for an in-depth discussion for the linear case). Thus we add this equation to the stationary state equations and solve for $b$ self-consistently. The bias $b$ itself  follows then implicitly some slow dynamics such that $\overline{g_t} = 0$ holds true. Since the value of $b$ has to be determined self-consistently we would not be able to advise  the agents  of how to appropriately adjust their bias in the choice of strategies right from the start.

For a linear price function $g$, the solution to the stationary state equations minimizes $q_A$ (from Eq.~\ref{eq: q_A}). The order parameter $q_A$ can be written as a spin glass Hamiltonian similar to that of the Hopfield model  \cite{challet_phase_1999}, which allowed \cite{challet_statistical_2000} to provide a solution to the linear case by the replica method.
Our solutions $x_i^\ast, g_\mu^\ast$ will be given in terms of the cavity field $z_\mu$ and the corresponding response terms $R_x$ and $R_g$ in Eqs.~\ref{eq: x_i_ast}-\ref{eq: b_equation}  below. There we will proceed in analogy to the solution of the Hopfield model by means of the cavity method as in \cite{shamir_thouless-anderson-palmer_2000,mezard_mean-field_2017}
albeit somewhat adapted: Our calculation for the minority game corresponds to a zero temperature version of the calculations of
\cite{shamir_thouless-anderson-palmer_2000,mezard_mean-field_2017}, where in addition $\{g_\mu\}$ and $\{x_i\}$ are real numbers rather than discrete spins. Note that we have to deal with $N$ agents and $P$ information items (signals) $\mu$ both representing the market. The agents are characterized by microscopic variables $x_i$, the information items by price functions $g_\mu$, depending via $A_\mu$ on $x_i$.

\section{Cavity method for the stationary state} \label{sec: linear_cavity}
\subsubsection{Mean field theory and the cavity method} \label{sec: CLT}
All quantities that occur in $A^t$ and $A_\mu$ are random variables due to the initial random drawing of the strategies, so for large $N$ and $P$ it is very tempting to use the central limit theorem (on which Eq.~\ref{eq: A_mu_gaussian} was based), especially as we saw that for some values of $\alpha$ the arbitrage $A^t$ follows a Gaussian distribution (Fig.~\ref{fig: histogram}a).
Some care is needed: the central limit theorem requires the variables that are summed over to be uncorrelated, while the variables that occur here in the sums are correlated. There is a large number of agents, so the response of an agent to any particular other agent is small, implying that also the correlations between the agents are small. Nevertheless, ignoring these correlations corresponds precisely to ignoring an agents' impact on the market (i.e., on the rest of the agents).
For example, if we want to calculate the variance $\langle (A^t + \eta^t - b)^2 \rangle_s$ (where the average is over the initial random drawing of the strategies), we cannot assume that any two agents are uncorrelated: $\langle a^t_i a^t_j \rangle_s \neq \langle a^t_i \rangle_s \langle a^t_j \rangle_s$. In cases in which the Curie-Weiss mean-field theory becomes exact, the correlations are of order $1/N$, and do not contribute to macroscopic observables for $N\rightarrow\infty$.

In the minority game, and in spin glasses in general, the correlations are (in the replica-symmetric phase) of order $1/\sqrt{N}$, and cannot be neglected \cite{mezard_spin_1986}. Indeed, capturing these correlations was the point of introducing the learning dynamics in the first place, as learning implies exploiting correlations, while ignoring them simply gives back the 'no anti-coordination' equilibrium where (in this case due to the random value of $\mu$) each agent chooses $a = \pm 1$ independently with probabilities $\frac{1}{2} \pm \frac{1}{2}b/\sqrt{N}$. The cavity method gives a way to correct the mean-field results for the correlations between agents, by assuming that the correlations are small enough to be treated within linear response. This allows the Curie-Weiss mean-field, which in this context becomes the 'cavity field', to be corrected by a linear 'reaction term' that captures the feedback of the market and the agents to lowest order. Setting the reaction term to zero, the cavity field reduces to the Curie-Weiss mean-field, otherwise it depends itself on the reaction term. The correlations induced by `the market' will be expressed in terms of two types of fluctuating entities to which a single agent is exposed: the other agents' decisions and the information items $\mu$.

\subsection{Cavity assumptions for the minority game} \label{sec: assumptions}
The cavity assumption states that any microscopic quantity from $\{x_i\}$ and $\{g_{\mu}\}$ depends, up to $O(1/N)$, only linearly on any of the other microscopic quantities (which we afterwards self-consistently verify). The assumptions are written in terms of the conditional values $x^{(i, \mu)}_j(x_i, g_{\mu})$ and $g^{(i,\mu)}_{\nu}(x_i, g_{\mu})$, denoting the values of $x_j$ and $g_\nu$ conditional on $x_i$ and $g_\mu$ taking given values. Here $j \neq i$, $\nu \neq \mu$, and the superscript $(i,\mu)$ indicates that the values depend on $x_i$ and $g_\mu$. The cavity assumptions then state:

\begin{alignat}{4} \label{eq: assumption_1}
    g^{(i,\mu)}_{\nu}(x_i, g_{\mu}) &= g^{(i,\mu)}_{\nu}(0, 0) \, &+ \,\frac{\partial g^{(i,\mu)}_{\nu}(x_i, g_{\mu})}{\partial x_i}\biggr \rvert_{0,0} \cdot x_i &+ \frac{\partial g^{(i, \mu)}_\nu(x_i, g_{\mu})}{\partial g_\mu}\biggr \rvert_{0,0}  \cdot g_{\mu} \,&+\, O(1/N) \,, \\
    x^{(i, \mu)}_j(x_i, g_{\mu}) &= x^{(i, \mu)}_j(0, 0) \, &+  \, \frac{\partial x^{(i, \mu)}_{j}(x_i, g_{\mu})}{\partial x_i}\biggr \rvert_{0,0}  \cdot x_i &+ \frac{\partial x^{(i, \mu)}_{j}(x_i,g_{\mu})}{\partial g_\mu}\biggr \rvert_{0,0} \cdot g_{\mu}  \,&+\, O(1/N) \,. \label{eq: assumption_2}
\end{alignat}
Here the derivatives are assumed to be of order $1/\sqrt{N}$.
\subsection{Self-consistent solution}
Inserting the cavity assumptions (Eqs.~\ref{eq: assumption_1}-\ref{eq: assumption_2}) into the stationary state equations, the Eqs.~\ref{eq: stat_eq_3}-\ref{eq: stat_eq_b} become up to $O(1/\sqrt{N})$:
\begin{align}
    \forall i:  \quad & R_x x_i - z_i  =
        \begin{cases}
            0  &\text{if} \quad -1 < x_i < 1 \,, \\
            > 0 & \text{if} \quad x_i = 1 \,,
            \\
            < 0 & \text{if} \quad x_i = -1 \,,
        \end{cases} \,, \label{eq: cavity_eq_x}\\
    \forall \mu:& \quad \qquad A_\mu = z_\mu + R_g g_\mu \label{eq: cavity_eq_A}\,,
\end{align}
here $z_i$ and $z_\mu$ are cavity fields and $R_x x_i$ and $R_g g_\mu$ are reaction terms, worked out in \ref{app: cavity} to be given by:
\begin{alignat}{1}
    z_{i} &= \frac{1}{\sqrt{N}}\sum_{\nu \neq \mu} \xi_i^\nu  g_{\nu}^{(i, \mu)}(0,0) + O(1/\sqrt{N})  \,, \label{eq: z_i_eq} \\
    z_{\mu} &= \frac{1}{\sqrt{N}}\omega_i^\mu + \frac{1}{\sqrt{N}}\sum_{j \neq i} \big(\omega_j^\mu + \xi_j^\mu x_{j}^{(i,\mu)}(0,0) \big) + O(1/\sqrt{N})\,, \label{eq: z_mu_eq}
\end{alignat}
and
\begin{alignat}{5}
    R_x &=& -&\frac{1}{\sqrt{N}}&\sum_{\nu \neq \mu} \xi_i^\nu \frac{\partial g^{(i, \mu)}_\nu(x_i, g_\mu)}{\partial x_i}\biggr \rvert_{0,0} &+ O(1/N) \,, \label{eq: R_x_eq}\\
    R_g &=& &\frac{1}{\sqrt{N}}&\sum_{j \neq i} \xi^\mu_j \frac{\partial x^{(i, \mu)}_{j}(x_i,g_{\mu})}{\partial g_\mu}\biggr \rvert_{0,0} &+ O(1/N) \,. \label{eq: R_g_eq}
\end{alignat}
Although the expressions on the right hand sides of Eqs.~\ref{eq: z_i_eq}-\ref{eq: R_g_eq} are given for the specific agent $i$ and signal $\mu$, the left hand sides are respectively independent of $\mu$ (Eq.~\ref{eq: z_i_eq}), independent of $i$ (Eq.~\ref{eq: z_mu_eq}), and independent of both $i$ and $\mu$ (Eqs.~\ref{eq: R_x_eq} and \ref{eq: R_g_eq}). This is justified in \ref{app: cavity}. \\
The cavity field $z_i$ ($z_\mu$) gives the solution of the stationary state equation for agent $i$ (signal $\mu$) if agent $i$ (signal $\mu$) would have no influence on the rest of the system, i.e. if we would set $x_i$ ($g_\mu$) equal to $0$. $R_x \leq 0$ ($R_g \leq 0$) quantifies the reaction of the market to agent $i$ (the reaction of the agents to signal $\mu$). 
We can write the solution to Eqs.~\ref{eq: cavity_eq_x}-\ref{eq: cavity_eq_A} as:
\begin{align}
    x^\ast_i &= \hat{x}(z_i) \, \label{eq: x_i_ast} \\
    g^\ast_\mu &= \hat{g}(z_\mu) \label{eq: g_mu_ast} \,
\end{align}
where
\begin{align}
    \hat{x}(z_x) &\equiv \begin{cases}
            \frac{z_x}{R_x} &\text{if} \quad -1 < \frac{z_x}{R_x} < 1 \,, \\
            1 & \text{if} \quad \frac{z_x}{R_x} \geq 1 \,,
            \\
            -1 & \text{if} \quad  \frac{z_x}{R_x} \leq -1 \,,
        \end{cases} \,. \label{eq: xhat} 
\end{align}
and
where the function $\hat{g}$ is defined implicitly by the equation
\begin{align}
    \hat{g}(z_g) = \langle g(z_g + R_{g} \hat{g}(z_g) + \delta + \eta) \rangle_{\delta, \eta} \,. \label{eq: ghat}
\end{align}
Here we have replaced the discrete indices $i$ and $\mu$ by continuous ones $x$ and $g$. The expressions hold for an arbitrary agent $i$ and an arbitrary signal $\mu$, which allows for a self-consistent calculation of the cavity fields and reactions terms. The full calculation is given in \ref{app: cavity}. The cavity fields $z_i$ and $z_\mu$ are sums of independent variables and are self-consistently determined to be Gaussian random variables with distributions:
\begin{alignat}{1}
    P(z_i) &= P(z_x) = N \big(0, q_g/2) \, \label{eq: P(z_x)}\\
    P(z_{\mu}) &= P(z_g) = N\big(b, 1/2 + q_x/2 \big) \,,\label{eq: P(z_g)}
\end{alignat}
with $q_x$ as in Eq.~\ref{eq: q_x} and  $q_g \equiv \frac{1}{P} \sum_{\mu} g_{\mu}^2$.
Since we have the values of $x_i$ and $g_\mu$ as a function of the cavity fields (Eqs.~\ref{eq: x_i_ast}-\ref{eq: g_mu_ast}), 
the distributions of the cavity fields $z_x$ and $z_g$ allow us to derive the macroscopic parameters $q_x$ and $q_g$ from the expressions for the agents $\{x_i\}$ and market $\{g_\mu\}$:
\begin{alignat}{3}
    q_g &= \frac{1}{P}\sum_{\mu} (g^\ast_\mu)^2 &= \frac{1}{P}\sum_{\mu} \hat{g}(z_\mu)^2 &=\int \mathrm{d}z_g \,\hat{g}(z_g)^2 P(z_g)\,, \label{eq: q_g_solution}\\
    q_x &=  \frac{1}{N} \sum_i  (x^\ast_i)^2 &= \frac{1}{N} \sum_i  \hat{x}(z_i)^2 &= \int \mathrm{d}z_x \,\hat{x}(z_x)^2 \, P(z_x) \,, \label{eq: q_x_solution}
\end{alignat}
Self-consistent expressions for the reaction terms are worked out in \ref{app: cavity} to be:
\begin{alignat}{3}
    R_x &=&-&\frac{\alpha}{2} \int \text{d}z_g \, \hat{g}'(z_g) P(z_g) \leq 0\,,\label{eq: R_x} \\
    R_g &=& &\frac{1}{2}  \int \text{d}z_x \, \hat{x}'(z_x) P(z_x) \leq 0\,. \label{eq: R_g}
\end{alignat}
Finally, to find the bias $b$ we set $\overline{g_t} = \frac{1}{P}\sum_{\mu} g^\ast_{\mu} = 0$:
\begin{align} \label{eq: b_equation}
    0 = \frac{1}{P}\sum_{\mu}  g^\ast_\mu  = \frac{1}{P}\sum_\mu \hat{g}(z_\mu) = \int \mathrm{d}z_g \hat{g}(z_g) P(z_g) \,.
\end{align}
The parameter $b$ enters the mean of $P(z_g)$. The distributions $P(z_x)$ and $P(z_g)$ are given in Eq.~\ref{eq: P(z_x)} and Eq.~\ref{eq: P(z_g)} in terms of $q_g$ and $q_x$ and can thus be found self-consistently.
For the special case of anti-symmetric $g$ the self-consistent equations agree with those found with generating functionals in \cite{coolen_minority_2005},  for the linear case the equations agree with those found by the replica method \cite{challet2000modelling, coolen_minority_2005}. We note also that a naive application of the Curie-Weiss mean-field theory corresponds to setting the reactions terms $R_x$ and $R_g$ to zero, which would give $\hat{x}(z_x) = \pm 1$ for any non-zero $z_x$, and hence would incorrectly conclude that all agents are frozen. \\
{} \\
\subsection{Self-consistent verification of the cavity assumptions and the phase transition}
Under the cavity assumptions, the self-consistent calculation (\ref{app: cavity}) gives:
\begin{equation} \label{eq: derivs}
    \begin{aligned}[c]
        \frac{\partial g^{(i,\mu)}_{\nu}(x_i, g_{\mu})}{\partial x_i}\biggr \rvert_{0,0} &= c_{i \nu} \cdot \hat{g}'(z_\nu) \\
        \frac{\partial x^{(i, \mu)}_{j}(x_i,g_{\mu})}{\partial g_\mu}\biggr \rvert_{0,0} &= c_{\mu j} \cdot\hat{x}'(z_j)
    \end{aligned}
    \qquad\qquad
    \begin{aligned}[c]
        \frac{\partial g^{(i, \mu)}_\nu(x_i, g_{\mu})}{\partial g_\mu}\biggr \rvert_{0,0} &= c_{\mu \nu} \cdot \hat{g}'(z_\nu) \\
        \frac{\partial x^{(i, \mu)}_{j}(x_i, g_{\mu})}{\partial x_i}\biggr \rvert_{0,0}  &= c_{ij} \cdot \hat{x}'(z_j)
    \end{aligned}
\end{equation}
where $c_{\mu j}$, $c_{i \nu}$, $c_{ij}$ and $c_{\mu \nu}$ are effective interaction coefficients of order $1/\sqrt{N}$, given up to $O(1/N)$ by (\ref{app: cavity}):
\begin{equation}
    \begin{aligned}[c]
        c_{i\nu} &= \frac{1}{\sqrt{N}} \xi_i^\nu + \frac{1}{\sqrt{N}}\sum_{j \neq i} \xi_j^\nu  \frac{\partial x^{(i, \mu)}_{j}(x_i, g_{\mu})}{\partial x_i}\biggr \rvert_{0,0} \,, \\
        c_{\mu j} &= \frac{1}{\sqrt{N}}\xi_j^\mu + \frac{1}{\sqrt{N}}\sum_{\nu \neq \mu} \xi_j^\nu \frac{\partial g^{(i, \mu)}_\nu(x_i, g_{\mu})}{\partial g_\mu}\biggr \rvert_{0,0} \,,
    \end{aligned}
    \qquad\qquad
    \begin{aligned}[c]
        c_{\mu \nu} &= \frac{1}{\sqrt{N}} \sum_{j \neq i} \xi_j^\nu \frac{\partial x^{(i, \mu)}_{j}(x_i,g_{\mu})}{\partial g_\mu}\biggr \rvert_{0,0} \,, \\
        c_{ij}  &= \frac{1}{\sqrt{N}} \sum_{\nu \neq \mu} \xi_j^\nu \frac{\partial g^{(i,\mu)}_{\nu}(x_i, g_{\mu})}{\partial x_i}\biggr \rvert_{0,0} \,.
    \end{aligned}
\end{equation}
The cavity assumptions state that all the partial derivatives of Eq.~\ref{eq: derivs} are of order $1/\sqrt{N}$, which holds as long as $\hat{g}'(z_g)$ and $\hat{x}'(z_x)$ stay finite.
With the definitions of $\hat{x}(z_x)$ and $\hat{g}(z_g)$ (Eqs.~\ref{eq: xhat} and \ref{eq: ghat}) we can explicitly evaluate the derivatives:
\begin{alignat}{1}
    \hat{x}'(z_x) &= \mathbbm{1}(-1 < \hat{x}(z_x) < 1)/R_x \label{eq: x_deriv} \\
    \hat{g}'(z_g) &= \frac{1}{1/\langle g'(z_g + R_g \hat{g}(z_g) + \delta + \eta)\rangle_{\delta, \eta} - R_g} \,.
\end{alignat}
$R_g$ is negative, and $g'$ is positive, so $\hat{g}'(z_g)$ always stays finite.
On the other hand, $\hat{x}'(z_x)$ diverges when $R_x \rightarrow 0$, so we should find out whether this can occur. By Eqs.~\ref{eq: R_g} and \ref{eq: x_deriv}, $R_x \rightarrow 0$, also gives $1/R_g \rightarrow 0$. Expanding around this limit, $\hat{g}(z_g)$ becomes linear in $z_g$:
\begin{align}
    \hat{g}(z_g) &= \frac{ (z_g - b)}{1/\langle g'(b + \delta + \eta)\rangle_{\delta, \eta} - R_g} + O\big(1/R_g^3\big) \,,
\end{align}
with the bias $b$ such that $\langle g(b + \delta + \eta)\rangle_{\delta, \eta} = 0$,
 as can be confirmed by inserting this expression in Eqs.~\ref{eq: ghat} and \ref{eq: b_equation} and expanding in $1/R_g$.
With this expression we can solve the Eqs.~\ref{eq: R_x}-\ref{eq: R_g} for $R_x$ and $R_g$ close to the phase transition:
\begin{align}
    \vspace{0.1cm} R_x &= \frac{1-\phi - \alpha}{2} \cdot \langle g'(b + \delta + \eta)\rangle_{\delta, \eta}  + O(1/R_g^2) \label{eq: R_x_2} \,, \\
    R_g &= \frac{\alpha}{1 - \phi - \alpha} \cdot \frac{1}{\langle g'(b + \delta + \eta)\rangle_{\delta, \eta}} + O(1/R_g^2) \label{eq: R_g_2} \,,
\end{align}
where we defined the fraction of frozen agents $\phi$ by:
\begin{alignat}{2}
    (1 - \phi) &= \frac{1}{N} \sum_i \mathbbm{1}(-1 < x_i < 1) &&= \int \mathrm{d}z_x \,\mathbbm{1}(-1 < \hat{x}(z_x) < 1) P(z_x).
\end{alignat}
$R_x$ goes to zero when $\alpha - (1-\phi) \to 0$, which implies $\hat{x}'(z_x) \rightarrow \infty$, and a breakdown of the cavity assumptions (which require that the partial derivatives of Eqs.~\ref{eq: derivs} stay of order $1/\sqrt{N}$). Solving the self-consistent equations (Eqs.~\ref{eq: xhat}-\ref{eq: b_equation}) numerically for the linear case gives $\alpha_c \approx 0.3374$.
In general, such a breakdown of the cavity assumptions is known to correspond to Replica Symmetry Breaking \cite{mezard_spin_1986, del_ferraro_cavity_2014}.
\footnote{We also refer to an insightful geometrical interpretation of the phase transition in the minority game given in \cite{chakraborti_statistical_2015}.}

 Concluding, the collective responses of the agents for different $i$ and of the market (for different signals $\mu$) give rise to effective interactions between the agents and the market. At the phase transition, $ \frac{\partial g^{(i,\mu)}_{\nu}(x_i, g_{\mu})}{\partial x_i}\biggr \rvert_{0,0} \rightarrow 0$, so the impact of an agent on the market is $0$: any effect of the agent on the market is counteracted by the response of other agents. At the same time, $q_g \rightarrow 0$, which means that $g_\mu$ is identically $0$ for all $\mu$. The agents are very reactive to the market: $\frac{\partial x^{(i, \mu)}_{j}(x_i,g_{\mu})}{\partial g_\mu}\biggr \rvert_{0,0} \rightarrow \infty$, which means that the agents adjust their strategies for any arbitrarily small opportunity $g_{\mu} \neq 0$ to make a profit. This divergence in the agents response to changes in the market also indicates that the cavity assumptions break down.

\subsection{Predictions of the self-consistent equations for the expected arbitrage for different price functions}\label{sec: noise}
\begin{figure}
    \centering
	\subfloat[]{\includegraphics[width = 0.45 \textwidth]{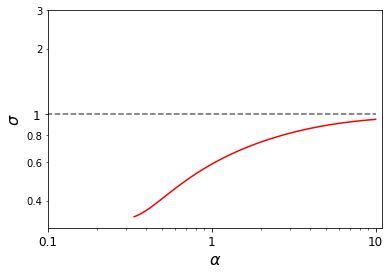}}
    \subfloat[]{\includegraphics[width = 0.45 \textwidth]{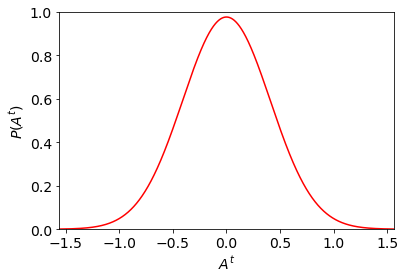}}
    \caption{(a) The volatility $\sigma$ (for a linear price function and no external noise) in the replica symmetric phase, as predicted by the self-consistent equations. (b) The distribution of $P(A^t)$ predicted by the self-consistent equations for $\alpha = 0.5$ (linear price function and no external noise), a Gaussian with width given by the volatility $\sigma$. The figures show the same results as those obtained by agent-based modeling (Figs.~\ref{fig: histogram}a and \ref{fig: different_S}b)}
    \label{fig: sigma_selfconsistent}
\end{figure}

The solution of the self-consistent equations (Eqs.~\ref{eq: xhat}-\ref{eq: b_equation}) yields the distributions of the cavity fields $z_x$ and $z_g$ (Eqs.~\ref{eq: P(z_x)}-\ref{eq: P(z_g)}), with which we can calculate macroscopic observables as in Eqs.~\ref{eq: q_g_solution}-\ref{eq: q_x_solution}. To calculate observables involving $A^\ast_\mu$, such as $q_A$ from Eq.~\ref{eq: q_A}, we can use the value of $A^\ast_\mu$ as a function of the cavity field $z_\mu$ (from Eq.~\ref{eq: cavity_eq_A}):
\begin{alignat}{1}
    A^\ast_\mu &= \hat{A}(z_\mu) \,, \\
    \hat{A}(z_g) &\equiv z_g + R_g \hat{g}(z_g) \,. \label{eq: Ahat}
\end{alignat}
As an example we use the self-consistent equations to rederive the results obtained by agent-based modeling in Sec.~\ref{sec: typical_behaviour}. By solving the self-consistent equations we plot the volatility (given by Eq.~\ref{eq: sigma}) as a function of $\alpha$, for a linear price function. This is shown in Fig.~\ref{fig: sigma_selfconsistent}a. Fig.~\ref{fig: sigma_selfconsistent}b shows the distribution of $P(A^t)$ for $\alpha = 0.5$. Figs.~\ref{fig: sigma_selfconsistent}a-b are in agreement with the results obtained by the direct agent-based modelling of the minority game given in Figs.~\ref{fig: histogram}a and \ref{fig: different_S}b. Linear and anti-symmetric price functions have been studied in-depth in \cite{challet2000modelling, coolen_minority_2005, papadopoulos_theory_2010}, where a variety of observables has been studied. Let us therefore focus specifically on price-functions that are not anti-symmetric, and focus on a phenomenon that is novel to the more general price function studied here: the subtle interplay between arbitrage and noise.

%
As mentioned in Sec.~\ref{sec: minority}, fluctuations can change the bias $b$ and therefore change the expected amount of arbitrage, as the expected amount of arbitrage is always equal to the bias:
\begin{alignat}{1}
    \overline{A^t + \eta^t} &= \frac{1}{P}\sum_\mu A_\mu = \int \text{d}z_g \hat{A}(z_g) P(z_g) = b \,
\end{alignat}
as the fluctuations of $\eta$ around the average are zero and $z_g$ was derived to be normally distributed around $b$.
At the same time the value of the bias should be such that:
\begin{align}
    \overline{g_t} = \int \mathrm{d}z_g \,\mathrm{d}\delta \, \mathrm{d}\eta \, g(\hat{A}(z_g) + \delta + \eta) \, P(z_g)P(\delta)P(\eta) = 0 \,. \label{eq: bias_eq}
\end{align}
The noise terms $\delta$ and $\eta$ have zero mean, so on average $\hat{A}(z_g) + \delta + \eta = b$. However, if $g$ is non-linear, the value of $b$ that solves Eq.~\ref{eq: bias_eq} may depend on the precise distribution of $\hat{A}(z_g) + \delta + \eta$ such that $0 = \overline{g_t} \neq g(b)$. A change in the value of $b$ changes the expected value of $A$, which in turn changes the distribution of internal noise according to Eqs.~\ref{eq: A_mu_gaussian} and \ref{eq: P(z_g)}. Non-linearity in the price function may thus combine with noise to give unexpected effects. For a general non-linear price function and a general distribution of noise these effects are difficult to quantify. Nevertheless, we can distinguish three different cases in which  general statements about the interplay between noise and nonlinearities are possible.

\subsubsection{No noise}
In the hypothetical case where the variance of both the external and internal noise is $0$ (i.e. $P(z_g)P(\delta)P(\eta)$ are delta functions), we have:
\begin{align}
    \int \mathrm{d}z_g \,\mathrm{d}\delta \, \mathrm{d}\eta \, g(\hat{A}(z_g) + \delta + \eta) \, P(z_g)P(\delta)P(\eta) &= g(\hat{A}(b)) = g(b) = 0 \,.
\end{align}
Thus the value of $b$ should be such that $g(b) = 0$.

\subsubsection{Anti-symmetric price function  around a constant}
If there is a value of $b$ around which $g(b + x) = -g(b - x)$ (anti-symmetric or linear $g$),
then from Eqs.~\ref{eq: ghat} and \ref{eq: Ahat} we have $\big[\hat{A}(b + (z_g - b)) - b \big] = - \big[\hat{A}(b - (z_g - b)) - b \big]$. This means that in the integration of Eq.~\ref{eq: bias_eq} the noise cancels out:
\begin{align}
    \int \mathrm{d}z_g \,\mathrm{d}\delta \, \mathrm{d}\eta \, g(\hat{A}(z_g) + \delta + \eta) \, P(z_g)P(\delta)P(\eta) & = g(b) = 0 \,.
\end{align}
Thus Eq.~\ref{eq: bias_eq} is solved for the $b$ around which $g(b + x) = -g(b - x)$. The bias $b$ is again such that $g(b) = 0$, and therefore the noise has no effect on the bias.

\subsubsection{Strictly convex ($g'' > 0$) or strictly concave ($g'' < 0$) price function}
\begin{figure}
    \centering
    \hspace{-0.45cm}

	\subfloat[]{\includegraphics[width = 0.45 \textwidth]{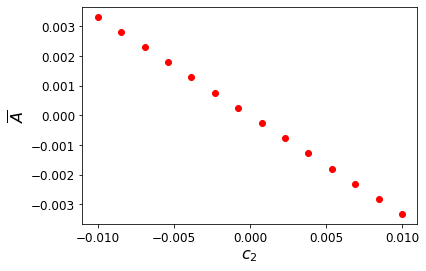}}
    \subfloat[]{\includegraphics[width = 0.45 \textwidth]{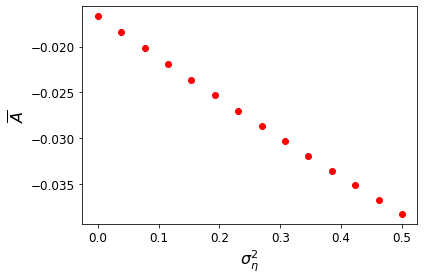}} \\
     \caption{(a) The expected amount of arbitrage, given by the bias, decreases with increasing second derivative of the price function, shown for the price function of Eq.~\ref{eq: nonlinear_g} and no external noise ($\sigma_\eta = 0$). (b) If the price function has positive second derivative, noise decreases the bias and thus the expected amount of arbitrage. Both panels are shown for the price function of Eq.~\ref{eq: nonlinear_g} with $c_2 = 0.05$, and for $\alpha = 1$.}
    \label{fig: noise_nonlinearity}
\end{figure}

If $g'' > 0$, then $g$ is strictly convex \footnote{This does not have to hold for the region where the distribution of $\hat{A}(z_g) + \delta + \eta$ is negligible, and so it is not in conflict with the assumption $g' > 0$.} and by Jensen's inequality \cite{boyd_convex_2004}:
\begin{alignat}{1}
    g(b) <& \int \mathrm{d}z_g \,\mathrm{d}\delta \, \mathrm{d}\eta \, g(\hat{A}(z_g) + \delta + \eta) \, P(z_g)P(\delta)P(\eta) = 0 \,, \label{eq: Jensen}
\end{alignat}
with the inequality reversed if $g$ is strictly concave ($g'' < 0 $).
For the left hand side to be smaller (larger) than $0$, the bias $b$, that is the expected amount of arbitrage, needs to be smaller (larger) than the value for which $g(b) = 0$. External noise in combination with a convex (concave) price function thus decreases the expected amount of arbitrage compared to the case where either the price function is anti-symmetric or where there is no noise. The larger (smaller) $g''$, or the stronger the noise, the larger the gap in the inequality of Eq.~\ref{eq: Jensen}, and thus the more the expected amount of arbitrage is changed. Note that these statements hold for an arbitrary distribution of both the external and internal noise, and for arbitrary convex (concave) price function $g$.
To illustrate this in a concrete example, we assume a small non-linearity in the price function:
\begin{align}
    g(x) = x + c_2 x^2 + 0.05 x^3 \,, \label{eq: nonlinear_g}
\end{align}
for some small constant $c_2$, and numerically solve the self-consistent equations (Eqs.~\ref{eq: xhat}-\ref{eq: b_equation}) for different values of $c_2$ and for different strengths of Gaussian external noise. The results are shown in  Figs~\ref{fig: noise_nonlinearity}a and b, respectively. Fig.~\ref{fig: noise_nonlinearity}a shows that increasing $c_2$ (and therefore $g''$) leads to a decrease in $\overline{A^t}=b$, where $b$ is the value of the bias for which the integral in Eq.~\ref{eq: Jensen} vanishes, while Fig.~\ref{fig: noise_nonlinearity}b shows that increasing the strength of the external noise (for $c_2 > 0$) decreases $b$ as well. \\

\section{Cavity method for the transient dynamics of the scores of strategies} \label{sec: dynamics}
In Sec.~2, we used the results of \cite{marsili_continuum_2001} to decompose the market volatility into three sources of noise (Sec.~\ref{sec: sources_of_fluctuations}) and to establish the stationary state equations (Sec.~\ref{sec: stat_eq}). This relies on the fact that for large times the minority game reaches a stationary state in which the agents' decisions $\{x_i^t  = \pm 1\}$ at time $t$ can be dynamically treated as   fluctuating independently up to $O(1/N)$ around the means $\{x_i\}$ (which themselves \textit{are} correlated on the order of $O(1/\sqrt{N})$ ). The authors of \cite{marsili_continuum_2001} derived their results for a modified version of the learning dynamics in which the short-time behaviour of the minority game is simplified, and conjecture that their results also hold for the unmodified version. Here we combine their approach with the cavity method. This allows us to derive the behaviour of how  the minority game changes depending on the timescale ($t \ll P$, $t \sim P$ and $t \gg P$). We show that the stationary state is reached due to an interaction between different sources of noise (both external noise $\eta$ and noise due to the random information parameterized by $\mu$), and the market impact of the agents, which on long time scales gives rise to fluctuations in the agents' decisions $\{x_i^t \}$. This ultimately confirms that the conclusions of \cite{marsili_continuum_2001} also hold for the unmodified learning dynamics of Sec.~\ref{sec: minority}. \\
The cavity method can be used to study the dynamics of mean-field spin glass systems \cite{mezard_spin_1986}.
As in the stationary state, one assumes that all microscopic quantities depend on each other only weakly, such that the dependencies can be treated as linear. To  investigate the dynamics of the strategy scores, we will only need the dependence of $g_t$ on a given agent $i$, rather than all four dependencies between the agents and the market (as in Sec.~\ref{sec: linear_cavity}) which would be required for self-consistently determining the time-dependent cavity and response terms. Dynamically, the quantity $g_t$ at time $t$ depends on an agent $i$ not only through the agent's decision $x_i^t = \text{sign}(U_i^t)$ at time $t$, but also through all agent $i$'s decisions $\{x_i^{t'}\}$ at times $t' \leq t$. Writing out the dependency of the quantity $g_t$ on agent $i$ thus gives a function $g_t^{(i)}(\{x_i^{t'}\})$ depending on all $\{x_i^{t'}\}$ for $t' \leq t$.
Within the cavity method, the effect of an agent on the rest of the system should then be treated within linear response theory  \cite{mezard_spin_1986}, by writing the update of the scores for agent $i$ as:
\begin{align} \label{eq: dynamical_cavity}
    U^t_i - U_i^{t-1} = - \xi_i^{\mu^t}g_t^{(i)}(\{x_i^{t'}\}) = - \strike{z}_t^i + \sum_{t' \leq t}r^i_{t,t'}x^{t'}_i \,,
\end{align}
with
\begin{align}
    \strike{z}_t^i &= \xi_i^{\mu^t}g_{t}^{(i)}(0)  \, \label{eq: h}\\
    r^i_{t,t'} &= -\xi_i^{\mu^t}\frac{\partial g_t^{(i)}(\{x_i^{t''}\})}{\partial x^{t'}_i}\biggr\rvert_{\{x^{t''}_i\} = 0 } \,, \label{eq: r}
\end{align}
The curly brackets represent all time instants $t''\le t$. The response to a change at time $t'\le t$ is evaluated at $\{x^{t''}_i\} = 0$ for all $t''$. The time evolution of the score $U_i^t$ depends on a cavity term $\strike{z}_t^i$, which is independent of the actions of the agent $i$, as it is evaluated at $\{x_i^{t''}\}=0$ and a response term $ \sum_{t' \leq t}r^i_{t,t'}x^{t'}_i$ which gives the impact of the agent's decision at all times $t' < t$ on the market price $g_{t}$ at time $t$. This means that the cavity ansatz is directly applied to the evolution of the price dynamics which enters the scores of the strategies. Comparing Eq.~\ref{eq: dynamical_cavity} with Eqs.~\ref{eq: stat_eq_3} and \ref{eq: cavity_eq_x}, we find that, in the replica-symmetric phase, $\overline{\strike{z}_t^i} = \frac{1}{\sqrt{\alpha P}}z_i$ and $\overline{\sum_{t' \leq t}r^i_{t,t'}x^{t'}_i} = \frac{1}{\sqrt{\alpha P}}R_x x_i$. The two terms of Eq.~\ref{eq: dynamical_cavity} thus correspond to dynamical versions of the cavity and reaction terms.
Since the value of $\strike{z}_t^i$ depends on $\mu^t$ and $\eta^t$, which are random variables, $\strike{z}_t^i$ itself also becomes a random variable.
Denoting an average over the distribution of $\mu^t$ and $\eta^t$ by $\langle \mathbbm{O}^t_{\mu}(\eta) \rangle_{\mu, \eta} \equiv \frac{1}{P} \sum_{\mu} \big( \int \mathrm{d} \eta P(\eta) \mathbbm{O}^t_{\mu}(\eta) \big)$, the $\strike{z}^i_t$ are random variables whose mean, autocorrelation, time correlations and agent-correlations, respectively,  are given by:
\begin{align}
    \langle \strike{z}^i_t \rangle_{\mu, \eta} &=  O(1/\sqrt{P}) \,,\\
    C_{t,t} \equiv \langle \strike{z}^i_t \strike{z}^i_{t} \rangle_{\mu, \eta} - \langle \strike{z}^i_t\rangle_{\mu, \eta} \langle \strike{z}^i_{t} \rangle_{\mu, \eta} &=  O(1) \,, \label{eq: C}\\
   \langle \strike{z}^i_t \strike{z}^i_{t'} \rangle_{\mu, \eta} - \langle \strike{z}^i_t\rangle_{\mu, \eta} \langle \strike{z}^i_{t'} \rangle_{\mu, \eta} &= O(1/P) \,, \\
    \langle \strike{z}^i_t \strike{z}^j_{t} \rangle_{\mu, \eta} - \langle \strike{z}^i_t\rangle_{\mu, \eta} \langle \strike{z}^j_{t} \rangle_{\mu, \eta} &= O(1/P) \,. \label{eq: agent_correlations}
\end{align}
The mean is of $O(1/\sqrt{P})$ because of the central limit theorem as applied to the summation over $\mu$.
The quantity $C_{t,t}$ is the variance of $\strike{z}_i^t$. It gives an effective noise due to the random values of the signal $\mu^t$ and external noise $\eta^t$ and is $O(1)$ because of the $ \langle \strike{z}^i_t \strike{z}^i_{t} \rangle_{\mu, \eta}$ term. The correlations $\langle \strike{z}^i_t \strike{z}^i_{t'} \rangle_{\mu, \eta}$ and $\langle \strike{z}^i_t \strike{z}^j_{t} \rangle_{\mu, \eta}$ are sums of $O(P^2)$ terms with means zero, for which application of the central limit theorem leads to the indicated $O(1/P)$.
As for the response term, $g_t = O(1)$ is the aggregate contribution of $N$ agents; the dynamical version of the cavity assumption is then that the total response of $g_{t}$ to the actions of agent $i$ is of $O(1/\sqrt{N})$ ($= O(1/\sqrt{P})$):
\begin{align} \label{eq: dynamics_assumption}
    \sum_{t' \leq t} r^i_{t,t'} = O(1/\sqrt{P}) \,.
\end{align}
This assumption is consistent with the fact that in the replica-symmetric phase we have $\overline{\sum_{t' \leq t}r^i_{t,t'}x^{t'}_i} =  \frac{1}{\sqrt{\alpha P}} R_x x_i$, being of $O(1/\sqrt{P})$. By definition of $A^t$ we furthermore have that $r^i_{t,t}$ is negative (as the slope of the price function is positive) and $\sim 1/\sqrt{N}$. 
For $t' < t$, $r^i_{t,t'}$ tends to be positive (since an increase in $A^{t'}$ will cause the other agents to change their strategies). On average, however, $\sum_{t' \leq t}r^i_{t,t'}$ is negative due to the contributions from $r^i_{t,t}<0$, since from Sec.~\ref{sec: linear_cavity} we have that $R_x \leq 0$ (Eq.\ref{eq: R_x}). The assumption of Eq.~\ref{eq: dynamics_assumption} also implies that in the replica-symmetric phase the market response $r^i_{t,t'}$ decays to zero for large enough $t - t'$ (since the amount of terms that is summed over grows without bound for large $t$). This assumption breaks down in the immediate vicinity of the phase transition already in the high-$\alpha$ phase. In addition and in general, for spin glasses it is known that in the replica-symmetry broken phase ergodicity is broken \cite{mezard_spin_1986}. In the context of the minority game, the interpretation is that in the replica-symmetry broken phase the total response given on the left-hand side of Eq.~\ref{eq: dynamics_assumption} diverges for large $t$, because a temporary change in the strategy used by the agent has a permanent impact on the market. 
Whereas in the replica-symmetric phase $r^i_{t,t'}$ decays exponentially for large $t -t'$, in the replica-symmetry broken phase the decay follows a power law in $t-t'$ \cite{mezard_spin_1986}. We note that in actual financial markets the rate at which the impact of traders' actions decay in time, and conditions under which the market responds linearly to traders, are both active topics of research \cite{cont_price_2010,bouchaud_how_2008}.

\subsection{Short timescales $t \ll P$: A random walk}
\begin{figure}
    \centering
	\subfloat[]{\includegraphics[width = 0.45 \textwidth]{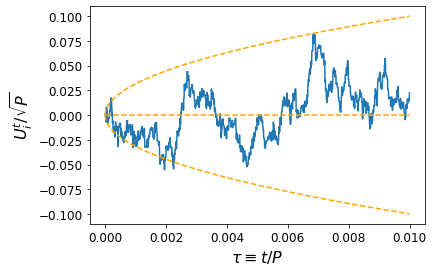}}
    \subfloat[]{\includegraphics[width = 0.45 \textwidth]{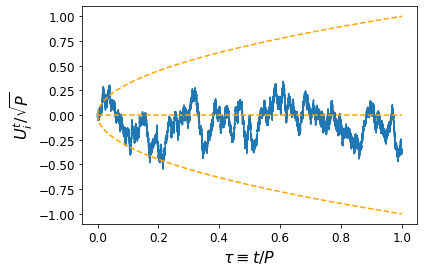}} \\
    \subfloat[]{\includegraphics[width = 0.45 \textwidth]{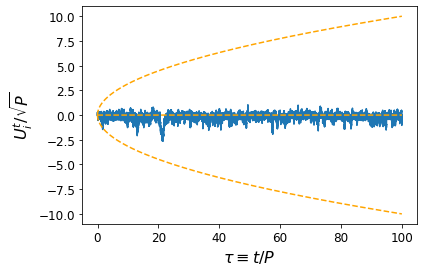}}
    \subfloat[]{\includegraphics[width = 0.45 \textwidth]{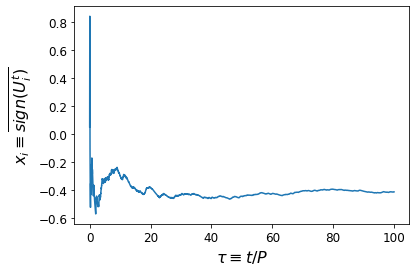}}
     \caption{The evolution of $U_i^t$ on (a) a timescale $t \ll P$ (b) a timescale $t \sim P$ and (c) a timescale $t \gg P$. The results shown are for a typical agent (the same agent is shown in (a)-(d)), linear price function $g$ and no noise, $N = 10^4$, $P = 2\cdot10^5$ and $S = 2$. The dotted lines, given by $\pm \sqrt{t}$, indicate the typical range of fluctuations of a random walk. (a) $U_i^t$ behaves as a random walk for $t \ll P$.
     (b) For $t \sim P$, $U_i^t$ stays smaller than  expected for a random walk, and instead fluctuates around $0$. (c) On timescales $t \gg P$ the individual fluctuations around $0$ cannot be resolved, such that $\text{sign}(U_i^t)$ effectively becomes binary noise characterized only by its mean value $\overline{\text{sign}(U_i^t)} \equiv x_i$. (d) Convergence of $x_i \equiv \overline{\text{sign}(U_i^t)}$ on a timescale of $t \gg P$.}
    \label{fig: timescales}
\end{figure}

The fluctuations in $\strike{z}_i^t$, with variance $C_{t,t}$, are due to the external noise and the random drawing of $\mu$. $C_{t,t}$ is of order $1$, while the mean of $\strike{z}_i^t$ and the response term are of order $1/\sqrt{P}$.
On short time scales the noise given by the fluctuations in $\strike{z}_i^t$ therefore dominates, and the scores effectively follow a random walk. $U_i^t$ is thus expected to grow $\sim \sqrt{t}$ on this timescale, as shown in Fig.~\ref{fig: timescales}a.
Note that this statement holds for the replica-symmetric phase, where the assumption of Eq.~\ref{eq: dynamics_assumption} holds;
without this assumption, the response term may become important already on shorter timescales.

\subsection{Timescale $t \sim P$: coarse graining over time}
\begin{figure}
    \centering
	\subfloat[]{\includegraphics[width = 0.45 \textwidth]{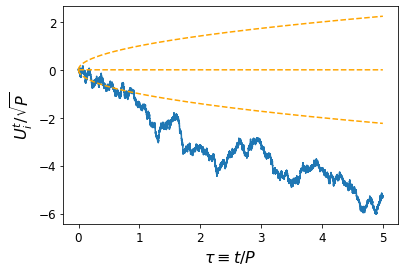}}
    \subfloat[]{\includegraphics[width = 0.45 \textwidth]{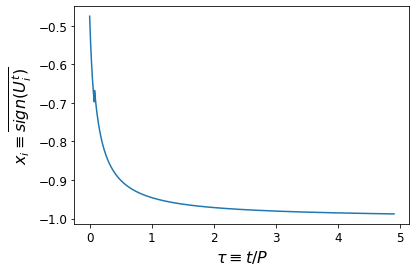}}
     \caption{(a) The evolution of $U_i^t$ and (b) the convergence of $x_i \equiv \overline{U_i^t}$ on a timescale of $t \sim P$, for a typical \textit{frozen} agent, linear price function $g$ and no noise, $N = 10^4$, $P = 2\cdot10^4$ and $S = 2$, showing that for frozen agents $|U_i^t|$ can grow without bound. In our formalism this in interpreted by the fact that for this agent the cavity term $z_i^t$ (of Eq.~\ref{eq: delta_tau_dynamics}) is generally larger than the response term, and hence the response term cannot drive $U_i^t$ back to $0$. The dotted lines are given by $\pm \sqrt{t}$.}
    \label{fig: frozen}
\end{figure}

On timescales $t \sim P$, the dynamics gets more interesting: the contribution of the noise due to $\eta$ and $\mu$ to $U_i^t$ is $O(\sqrt{P})$, while the contributions of the mean of $\strike{z}$ and of the response term are of order $\frac{P}{\sqrt{P}} = \sqrt{P}$ and hence of the same order as the contribution of this noise.
On these timescales the different terms therefore interact, and the scores follow more intricate dynamics than a simple random walk. Since as long as $\Delta t \ll P$ the scores follow simply a random walk, let us coarse-grain the dynamics in time (following \cite{marsili_continuum_2001} with respect to the coarse-graining), looking only at the changes in score every $\Delta t$-th time step. Over this time interval the decisions of the agents are still uncorrelated, and the central limit theorem applies when we sum over $\Delta t=P \cdot d\tau$ time steps, being $\propto P\gg 1$ steps, while $d\tau$ is a small time interval that will later be treated as infinitesimally small. Thus we rescale the time as $\tau = t / P$:
\begin{align}
    \tau &= t / P \,,\\
    \text{d} \tau &= \Delta t / P \,.
\end{align}
Over time intervals $\Delta t \propto P$ we have $U^t_i \propto \sqrt{P}$, so it is useful to look at the rescaled quantity $u_i^\tau \equiv U^t_i/\sqrt{P}$:
\begin{align}
	u_i^\tau &\equiv U^t_i/\sqrt{P} \\
	\text{d} u^\tau_i &= U^t_i/\sqrt{P} - U_i^{t-\Delta t}/\sqrt{P}.
\end{align}
To find the change in scores over $\Delta t$ time steps (i.e., $\text{d} u^\tau_i$), we can aggregate the terms in Eq.~\ref{eq: dynamical_cavity} by calculating the running averages (with $0 \leq t'' \leq t $, and $r_{t'',t'} = 0$ if $t'  > t''$):
\begin{align}
    R^i_{\tau, \tau'} &= \frac{1}{\text{d} \tau}\sum_{t'' = t - \Delta t +1}^{t} r^i_{t'', t'}/\sqrt{P} \,, \label{eq: R}\\
    z_\tau^i &= \frac{1}{\text{d} \tau}\sum_{t'' =  t - \Delta t + 1}^{t} \langle \strike{z}_{t''}^i \rangle_{\mu, \eta} /\sqrt{P}\,, \label{eq: z}\\
    \zeta_i^\tau &= \frac{1}{\text{d} \tau}\sum_{t'' =  t - \Delta t + 1}^{t} \big(\strike{z}_{t''}^i - \langle \strike{z}_{t''}^i \rangle_{\mu, \eta}\big) /\sqrt{P}  \sim N(0, \frac{1}{(\text{d} \tau)^2}\sum_{t'' =  t - \Delta t + 1}^t C_{t'',t''}/P)\,. \label{eq: zeta}
\end{align}
Here the rescaling by $1/\sqrt{P}$ is to keep the three quantities of $O(1)$.
The sums over time in Eqs.~\ref{eq: R}-\ref{eq: zeta} contain $O(P)$ terms. The individual terms in the sums of Eqs.~\ref{eq: R}-\ref{eq: z} are of $O(1/P)$, while the fluctuations in Eq.~\ref{eq: zeta} are about zero and therefore, by the central limit theorem, the whole term remains $O(1)$.
With these averaged quantities we obtain the dynamics:
\begin{align} \label{eq: delta_tau_dynamics}
    \frac{\text{d}u^\tau_i}{\text{d}\tau} &= z_\tau^i + \sum_{\tau' \leq \tau}R^i_{\tau,\tau'}\text{sign}(u^{\tau'}_i) + \zeta_i^\tau  \,,
\end{align}
where we explicitly inserted for $x_i^t \equiv \text{sign}(U_i^t) = \text{sign}(u_i^\tau)$. For $\text{d}\tau = \frac{\Delta t}{P} \rightarrow 0$, Eq.~\ref{eq: delta_tau_dynamics} corresponds to a stochastic differential equation, 
valid for timescales of $t \sim P$ and longer. The equation gives insight into the dynamics of $u_i^\tau = U_i^t/\sqrt{P}$. Note the non-Markovian dynamics of the second term. In comparison to the Langevin equation derived in \cite{marsili_continuum_2001}, the derivation and the meaning of the deterministic and the stochastic terms here is different.  In this derivation, the noise term $\zeta_i^\tau $ is due to price fluctuations caused by $\eta$ and $\mu$, derived as a Gaussian.  Which of the strategies is preferred by the agents is determined by the cavity term  $z_\tau^i$;
if on average it is positive it tends to drive $U_i^t$ upwards, and vice-versa. On the other hand, the response $R^i_{\tau,\tau'}$ tends to be negative, so if the cavity term  $z_\tau^i$ is not too strong, the response term tends to bring $U_i^t$ back to zero. The resulting behaviour is shown in Fig.~\ref{fig: timescales}b, where it can be seen that $U_i^t$ stays closer to $0$ than would be expected for a random walk. In terms of the resulting actions of the agent, $U_i^t$ fluctuating around zero means that $\text{sign}(U_i^t)$ fluctuates between $\pm 1$, thus creating a balance between the different strategies that are used by the agent.
Here it should be noticed that a zoom into short time scales of Fig.~\ref{fig: timescales}b of the order of $t=P/\tau=0.010$ would again show a time evolution of a random walk as in Fig.~\ref{fig: timescales}a, the random walk is not restricted to the transient time until the stationary state is reached.
On the other hand, if the cavity term remains stronger than the response term, it drives $U_i^t$ away from zero, resulting in a 'frozen' agent that never changes its strategy. This is shown in Fig.~\ref{fig: frozen}.

Although not shown here, it is generally known \cite{challet2000modelling, coolen_minority_2005} that for $\alpha > \alpha_c$ an increase in $\alpha$ leads to a decrease in the fraction of frozen agents, as can be found either by agent-based modelling or by a numerical solution of the self-consistent equation. In Sec.~\ref{sec: linear_cavity} we showed that near the phase transition the fraction of frozen agents equals $\phi = 1 - \alpha_c \approx  0.6626$. 
\\
Now we can also verify the assumption of Sec.~\ref{sec: sources_of_fluctuations} about the dynamical correlations between different agents, which we used to write the volatility and the steady state equations in terms of the frequencies $\{x_i\}$. The assumption was $\overline{(x_i^t - x_i)(x_j^t - x_j)} = \overline{(x_i^t - x_i)} \cdot \overline{(x_j^t - x_j)}$.
The fluctuations in $x_i^t - {x_i}$ are due to the noise term $\zeta^{\tau'}_i$, and thus the correlations $\overline{(x_i^t - x_i)(x_j^t - x_j)}$ are due to the correlations $\overline{\zeta^{\tau}_i \zeta^{\tau}_{j}}$. Since, by Eqs.~\ref{eq: agent_correlations} and \ref{eq: zeta} we have that $ \overline{\zeta^\tau_i\zeta^\tau_{j}}$ is an order of $P$ smaller than $\overline{(\zeta^\tau_i)^2} $, we expect $\overline{(x_i^t - x_i)(x_j^t - x_j)} - \overline{(x_i^t - x_i)} \cdot \overline{(x_j^t - x_j)} \sim O(1/P)$,
 and thus the fact that we neglected them in Sec.~\ref{sec: sources_of_fluctuations} is justified.\\

\subsection{Long timescale: $t\gg P$, the stationary state}
On timescales $t \gg P$ the score $U_i^t$ keeps fluctuating around zero, but the individual fluctuations cannot be resolved on this timescale in a plot like Fig.~\ref{fig: timescales}c. The correlation time of these fluctuations is much smaller than the timescale, thus $\text{sign}(U_i^t) = \pm 1$ effectively becomes binary noise, randomly fluctuating between $\pm 1$ as characterized by its mean $\overline{\text{sign}(U_i^t)} \equiv x_i$ (shown in Fig.~\ref{fig: timescales}d) and resulting variance $\overline{\text{sign}(U_i^t)^2} - x_i^2 = 1 - x_i^2$. On these timescales the system thus reaches a stationary state, characterized by the stationary state equations as analyzed in Sec.~\ref{sec: linear_cavity}, which can be solved to find $x_i$. Indeed, in the long time limit we get exactly $\overline{z^i_\tau} = \frac{1}{\sqrt{\alpha}}z_i$ and $\overline{\sum_{\tau' \leq \tau} R_{\tau,\tau'}} = \frac{1}{\sqrt{\alpha}}R_x$,
corresponding directly to the cavity field and the reaction term in the stationary state.
$\frac{1}{\sqrt{\alpha}}R_x$ therefore also quantifies the error incurred by the agents themselves when updating the scores according to (Eq.~\ref{eq: score_definition}), in which the agents neglect their own impact on the market. Interestingly, we saw in Sec.~\ref{sec: linear_cavity} that approaching the phase transition $R_x \rightarrow 0$, so the error incurred by an individual agent for neglecting the response term becomes negligible.
Nevertheless, for the whole system the response term still plays a fundamental role \cite{marsili_exact_2000}. \\
The dynamical behaviour on the three timescales $t \ll P$, $t \sim P$ and $t \gg P$ is summarized in table~\ref{tab: timescales}.
On timescales $t \ll P$, the evolution of the scores of strategies is completely driven by noise (parameterized by $\eta$ and $\mu$).
On timescales $t \sim P$, the noise $\zeta_t^s$ is of the same order of magnitude as the cavity term $z_t^i$ and the response term $R^i_{t,t'}$. The scores fluctuate around $0$; on timescales $t \gg P$ 
$x_i^t = \text{sign}(U_i^t)$ effectively becomes binary noise characterized only by its mean $\overline{x_i^t} = x_i$. $x_i$ can then be found from the stationary state equations analyzed in Sec.~\ref{sec: linear_cavity}.
\begin{table}
    {\renewcommand{\arraystretch}{1.2}
    \begin{tabular}{c c c}
        \hline
        Timescale & Relevant variable & Dynamical behaviour\\
        \hline
        $t \ll P$ &  $U_i^t$ & \begin{tabular}{c}Random walk \\ (Fig.~\ref{fig: timescales}a) \end{tabular} \\
        \hline
        $t \sim P$ & $u_i^\tau \equiv U_i^t/\sqrt{P}$ & \begin{tabular}{c} Non-Markovian Stochastic Differential Equation \\ (Eq.~\ref{eq: delta_tau_dynamics}, Fig.~\ref{fig: timescales}b) \end{tabular} \\
        \hline
        $t \gg P$ & $x_i^t \equiv \text{sign}(U_i^t)$ & \begin{tabular}{c} Binary noise with $p(\text{sign}(U_i^t) = \pm 1) 
        \equiv \frac{1}{2} \pm \frac{1}{2}x_i $ \\ (Figs.~\ref{fig: timescales}c-d) \end{tabular}\\
        \hline
    \end{tabular}}
\caption{Dynamical behaviour of the scores on different timescales}
\label{tab: timescales}
\end{table}

\section{Conclusions and outlook}
The model that we used for the minority game is  minimal and stylized in view of financial markets. It is particularly stylized in view of the incorporation of the complexity of information provided by the market and of the number of possible strategies that the agents can use to exploit this information, modelled by single parameters $P$ and $S$, respectively. In this paper $S$ was fixed to two for simplicity. Besides $P$ and $S$, model parameters are $N,  P(\eta)$ and $g$. Here $N$ was the number of traders who behave as arbitrageurs and $P(\eta)$ described external, white noise due to trades of assets unrelated to arbitrage. In our previous work \cite{ritmeester_minority_2021} the application was to the energy market with $g$ the price function on the reserve energy market. The asset was the traded energy. More generally, on financial markets $g$ may represent the functional dependence of financial returns on excess demand of the traded assets.

The model has a counterpart to replica symmetry breaking at a critical value of the control parameter $\alpha=N/P=\alpha_c$. For $\alpha<\alpha_c$, the agent-based modeling of this minority game reproduces strongly non-Gaussian and clustered fluctuations in arbitrage, clustered in their size and in time, known to be present in financial markets \cite{cont_empirical_2001}. 

In combination with the cavity method, a study of the low-$\alpha$ phase is beyond the scope of this paper, but a number of insights were gained from applying the cavity method in the high-$\alpha$ phase with $\alpha>\alpha_c$. (i) The Gaussian distribution of the arbitrage and the volatility (variance of the market due to arbitrage) as a function of the control parameter, both obtained  via agent-based modelling, were analytically reproduced as a function of the model parameters $N,P,S,g$. Macroscopic quantities like $\sigma$ (the volatility), and the order parameters $q_x$ (a measure for a lack of diversity of strategies) and $q_A$ (the volatility due to varying external information $\mu$) were determined in terms of microscopic quantities $x_i$ (the preference between strategies of agent $i$), and $g_\mu$ (the expected price for fixed external information $\mu^t=\mu$).

(ii) We have  established a concrete relation between the market volatility and specific features of the price function based on a detailed understanding of the interaction of volatility with the nonlinearities of a general price function, not considered in this generality before. Strictly convex (concave) price functions guarantee a decrease (increase) of arbitrage with increasing volatility, respectively. What entered these predictions was the cavity field $z_g$ that was self-consistently determined from the solution of the stationary equations for the variables $x_i$, characterizing the agents, and $g_\mu$, characterizing the market. The information contained in the cavity field goes beyond a standard mean-field (on the level of the Curie-Weiss model), it depends on the feedback of the agents and the market to individual actions of agents and market. Insights on which features of the price function are essential for reducing the arbitrage, when external stochastic events increase the volatility,  may be used for proposing suitable price policies. Vice-versa, when a decrease in arbitrage after an increase in the volatility is observed in the trading behaviour that is modelled as a minority game, such an observation may be explained by a convex price function.

(iii) Furthermore, in contrast to the naive mean-field theory, the feedback of the market and the agents, which is included in the cavity method, reproduced the fact that the agents need not get frozen to a specific strategy, but also in the stationary state keep on changing their strategy with a possible preference for a subset (in our case for one of the two possible strategies).

(iv) Stylized facts that are observed in time-series of asset price, such as volatility clustering, heavy tails and aggregational gaussianity \cite{cont_empirical_2001} are known to arise particularly in the critical regime of minority games \cite{challet_minority_2001, challet_minority_2001-1}. From a general physics perspective this is not surprising, as critical behavior (in the vicinity of a singularity)   often shares universal features which can be captured by stylized models. It is thus instructive to zoom into the critical region close to the phase transition point where the cavity method breaks down as the conditions for its applicability get violated. From general spin-glass theory this is known to correspond to a loss of ergodicity. For the minority game we can read off from the formulas that close to the transition point the response of the market to an individual agent goes to zero, while the response of an agent to the market diverges. This may reflect an actual mechanism which indicates nearby strong critical fluctuations of the market, precursors of a drastic change: The traders are very sensitive to information and respond very actively, but so much that their impact on the market gets lost.

(v) To pursue the transient time evolution of the scores (on which agents base their decision between strategies) towards the stationary state and to zoom into the time scales within the stationary state, we applied the cavity method also to the response of the price function to all decisions of an agent at earlier times than the considered instant. In this case it was possible to pursue the competition of the three sources of noise: The external noise, the internal fluctuations due to the provided information (parameterized by $\mu$), and the strategy changes of the agents, based on scores in hindsight. On short time scales ($t\ll P$), the scores ($U^t_i/\sqrt{P}$) perform a random walk with step size of $O(1)$, where correlations in subsequent updates are negligible. On an intermediate time scale $t\sim P$, the scores evolve differently from a random walk, where the response terms have built up and prevent too large excursions from zero, effectively memorizing earlier decisions. For long times until the stationary state is reached ($t\gg P$) or for long time intervals within the stationary state, the time correlations in $U_i^t$ become negligible. The sign of $U_i^t$ is then effectively described by random binary noise, fluctuating between $+1$ and $-1$, but not necessarily uniformly (as quantified by $x_i$); thus there may be a bias in favor of one of the strategies without getting frozen to a single one.

(vi) The distributions of the cavity fields $z_i,z_\mu$ ($z_x,z_g$) and $z^{i}_\tau$ were all Gaussians as the result of applying the central limit theorem to sums of uncorrelated variables, which cannot be applied to sums of the $x_i$ and $A_\mu$ variables (which are correlated to $O(\sqrt{N})$); in the cavity method these correlations are split off to the response terms. Non-Gaussian features are expected in the low-$\alpha$ phase also for the stationary behavior and in the critical region of the phase transition, where the cavity method breaks down.

Future work may extend this investigation to the low-$\alpha$ phase of the minority game, in which the market impact is expected to play an even more pronounced role. From corresponding results in spin glass theory it is known that in the replica-symmetry-broken phase, a temporary change in the market may have a permanent impact on the strategies used by the agents and vice-versa, for which linear response is no longer an adequate description. Although an analysis of the replica-symmetry-broken phase is substantially more difficult than that of the replica-symmetric (high-$\alpha$) phase, extensions of the cavity method  may be used also in the replica-symmetry-broken phase \cite{mezard_spin_1986}. Beyond the minority game, a number of other heterogenous agent-models have been mapped to spin glass problems \cite{chakraborti_statistical_2015, martino_statistical_2006}. As such the cavity method may be applied to gain deeper insight also into these models.


\section*{Acknowledgment}
We  thank the Bundesministerium f\"ur Bildung und Forschung (BMBF) (grant number 03EK3055D) for financial support.\\
\section*{References}
\bibliographystyle{iopart-num}
\bibliography{main}

\appendix

\section{Details of the cavity calculations for the stationary state} \label{app: cavity}
We consider a slight generalization of the stationary state equations of Sec.~\ref{sec: stat_eq}:
\begin{alignat}{1}
    &\, \qquad \qquad \quad  \! \!  \frac{1}{P} \sum_{\mu} g_{\mu} = 0  \\
    \forall i: \quad & -\frac{1}{\sqrt{N}}\sum_{\mu} \xi_i^\mu g_{\mu} - \epsilon_i =
    \begin{cases}
        0  &\text{if} \quad -1 < x_i < 1 \,, \\
        > 0 & \text{if} \quad x_i = 1 \,,
        \\
        < 0 & \text{if} \quad x_i = -1 \,.
    \end{cases} \label{eq: stat_eq_eps_i}\\
    \forall \mu: \quad & \qquad \qquad \qquad \quad \: \: g_{\mu} = \langle g(A_{\mu} +  \delta + \eta) \rangle_{\delta, \eta} \,,  \\
   & \qquad \qquad \qquad \quad \: \: A_{\mu} = \frac{1}{\sqrt{N}}\sum_i (\omega_i^\mu + \xi_i^\mu x_i) + \epsilon_\mu\,.
\end{alignat}
Here $\epsilon_i$ and $\epsilon_\mu$ are arbitrary small perturbations of order $1/\sqrt{N}$. Setting $\epsilon_i, \epsilon_\mu \rightarrow 0$, we would recover the stationary state equations of Sec.~\ref{sec: stat_eq}. As discussed in the main text, the cavity method uses the conditional values $x_j^{(i,\mu)}(x_i, g_\mu)$ and $g_\nu^{(i,\mu)}(x_i, g_\mu)$. To make the definitions these objects precise, we need to specify the sense in which we are allowed to vary $x_i$ (for some specific agent $i$) and $g_{\mu}$ (for some specific signal $\mu$),
to find the dependence of all other variables on $x_i$ and $g_\mu$. Eq.~\ref{eq: stat_eq_3} and Eq.~\ref{eq: stat_eq_g} give $N + P$ stationary state equations, one for each agent and one for each signal.
If we would solve all $N + P$ equations, we would get a solution $\{x^\ast_j: j = 1, \dots, N, g^\ast_\mu: \mu = 1, \dots, P\}$ for all $N$ agents and all $P$ signals.
Instead we single out one agent, agent $i$, and one signal, signal $\mu$.
To find the reaction of the system to an agent $i$ for a given $i$ and to the market $g_\mu$ for a given $\mu$ we need to suppose, for the moment,
that agent $i$ and signal $\mu$ do not necessarily behave according to the stationary state equations
(Eqs.~\ref{eq: stat_eq_3}-\ref{eq: stat_eq_g}),
but rather have some arbitrary values $x_i$ and $g_\mu$.
Given that all other agents and signals do behave according to the stationary state equations,
we can find the values of $x_{j \neq i}$ and values $g_{\nu \neq \mu}$ that solve the remaining $(N-1) + (P-1)$ equations.
Thus the solutions are conditional on the values of $x_i$ and $g_\mu$,
so that we can write this solution as $\{x_j^{(i,\mu)}(x_i, g_\mu), g^{(i, \mu)}_\nu(x_i, g_\mu)\}$ (we assume self-consistently that for given $x_i$ and $g_\mu$ the remaining $(N-1) + (P-1)$ equations have a unique solution).
Given this solution, finding $x_i^\ast$ and $g_\mu^\ast$ simply requires solving the remaining equations:
\begin{align}
    -\epsilon_i -\frac{1}{\sqrt{N}}\xi_i^\mu g_\mu &-\frac{1}{\sqrt{N}}\sum_{\nu \neq \mu} \xi_i^\nu g^{(i, \mu)}_\nu(x_i, g_\mu) =
    \begin{cases}
        0  &\text{if} \quad 0 < x_i < 1 \,, \\
        > 0 & \text{if} \quad x_i = 1 \,,  \\
        < 0 & \text{if} \quad x_i = -1 \,,
    \end{cases} \,, \label{eq: remain_eq_x_i}\\
    g_\mu &= \langle g(A_\mu + \delta + \eta)\rangle_{\delta, \eta} \,,\\
    A_\mu &= \epsilon_\mu +  \frac{1}{\sqrt{N}}\Big(\omega_i^\mu + \xi_i^\mu x_i + \sum_{j \neq i} (\omega_j^\mu + \xi_j^\mu x_j^{(i,\mu)}(x_i,g_\mu))\Big)  \,. \label{eq: remain_eq_A_mu}
\end{align}
We use the cavity assumptions of Eqs.~\ref{eq: assumption_1}-\ref{eq: assumption_2} for $x_j^{(i, \mu)}(x_i, g_\mu)$ and $g_\nu^{(i,\mu)}(x_i, g_\mu)$, from which the following special cases are implied:
\begin{equation}\label{eq: x_imu}
    \begin{aligned}[c]
        g_{\nu}^{(i)}(x_i) &= g_{\nu \setminus i} + \Delta^\nu_i \cdot x_i \,, \\
        x_{j}^{(i)}(x_i) &= x_{j \setminus i} + \Delta^j_i \cdot x_i  \,,
    \end{aligned}
    \qquad\qquad
    \begin{aligned}[c]
        g_{\nu}^{(\mu)}(g_\mu) &= g_{\nu \setminus \mu} + \Delta^\nu_\mu \cdot g_\mu \,, \\
        x_j^{(\mu)}(g_{\mu}) &= x_{j \setminus \mu} + \Delta^j_{\mu} \cdot g_{\mu} \,,
    \end{aligned}
\end{equation}
where we defined the abbreviations:
\begin{equation}
    \begin{aligned}[c]
        g_{\nu \setminus i} &\equiv g_\nu^{(i)}(0) \,,\\
        x_{j \setminus i} &\equiv x_j^{(i)}(0) \,,
    \end{aligned}
    \qquad\qquad
    \begin{aligned}[c]
        g_{\nu \setminus \mu} &\equiv g_\nu^{(\mu)}(0) \,, \\
        x_{j \setminus \mu} &\equiv x_j^{(\mu)}(0)  \,,
    \end{aligned}
\end{equation}
and:
\begin{equation} \label{eq: Delta_def}
    \begin{aligned}[c]
        \Delta^\nu_i &\equiv \frac{\partial g^{(i)}_{\nu}(x_i)}{\partial x_i}\biggr \rvert_{0} \,, \\
        \Delta^j_i  &\equiv \frac{\partial x^{(i)}_j(x_i)}{\partial x_j}\biggr \rvert_{0} \,,
    \end{aligned}
    \qquad\qquad
    \begin{aligned}[c]
        \Delta^\nu_\mu &\equiv\frac{\partial g^{(\mu)}_{\nu}(g_\mu)}{\partial g_\mu}\biggr \rvert_{0} \,,\\
        \Delta^j_\mu &\equiv \frac{\partial x^{(\mu)}_j(g_\mu)}{\partial g_\mu}\biggr \rvert_{0} \,.
    \end{aligned}
\end{equation}
With the same abbreviations we can write the original cavity assumptions of Eqs.~\ref{eq: assumption_1}-\ref{eq: assumption_2} as:
\begin{alignat}{4}
    g^{(i,\mu)}_{\nu}(x_i, g_{\mu}) &= g_{\nu \setminus i, \mu} \, &+ \,  \Delta^\nu_i \cdot x_i &+ \Delta^\nu_\mu  \cdot g_{\mu} &+ O(1/N) \,, \label{eq: cavity_abbrv_1} \\
    x^{(i, \mu)}_j(x_i, g_{\mu}) &= x_{j\setminus i,\mu} \, &+ \,\Delta^j_i  \cdot x_i &+ \Delta^j_\mu \cdot g_{\mu}  &+ O(1/N) \,. \label{eq: cavity_abbrv_2}
\end{alignat}
We used that the derivatives of Eqs.~\ref{eq: cavity_abbrv_1}-\ref{eq: cavity_abbrv_2} are, up to $O(1/N)$, equal to those of Eq.~\ref{eq: Delta_def}: We have $\frac{\partial g^{(i)}_\nu(x_i)}{\partial x_i}\biggr \rvert_{0} =\frac{\partial g^{(i, \mu)}_\nu(x_i, g^{(i)}_{\mu}(x_i))}{\partial x_i}\biggr \rvert_{0,g_\mu^{(i)}(0)} = \frac{\partial g^{(i, \mu)}_\nu(x_i, g_\mu)}{\partial x_i}\biggr \rvert_{0,0} + O(1/N)$, and equivalently for the other derivatives.
\subsection{The cavity field and the reaction term} \label{sec: calculation1}

\subsubsection{Equation for the agents to determine $x_i^*$}

Entering the assumption of Eq.~\ref{eq: cavity_abbrv_1} into the Eq.~\ref{eq: remain_eq_x_i} for $x_i$, we directly obtain:
\begin{align} \label{eq: cavity_eq_x_app}
R_x x_i - (z_{i \setminus \mu} + c_{\mu i} g_{\mu} + \epsilon_i) &=
    \begin{cases}
        0  &\text{if} \quad -1 < x_i < 1 \,, \\
        > 0 & \text{if} \quad x_i = 1 \,,
        \\
        < 0 & \text{if} \quad x_i = -1 \,,
    \end{cases} \\
z_{i \setminus \mu} &\equiv \frac{1}{\sqrt{N}}\sum_{\nu \neq \mu} \xi_i^\nu  g_{\nu \setminus i, \mu} \,, \label{eq: zi/mu} \\
c_{\mu i} &\equiv \frac{1}{\sqrt{N}}\xi_i^\mu + \frac{1}{\sqrt{N}}\sum_{\nu \neq \mu} \xi_i^\nu \Delta_{\mu}^\nu \,, \label{eq: c_mui}\\
R_x &\equiv -\frac{1}{\sqrt{N}}\sum_{\nu} \xi_i^\nu \Delta^\nu_i + O(1/N) \\
&= -\frac{1}{\sqrt{N}}\sum_{\nu} \langle \xi_i^\nu \Delta^\nu_i \rangle_s + O(1/\sqrt{N}) \,,
\end{align}
where in the last line we used the central limit theorem, as the terms for different $\nu$ are uncorrelated up to $O(1/N)$ \footnote{For signals $\nu$ and $\lambda$, we have $\langle \xi_i^\nu \Delta^\nu_i \cdot \xi_i^\lambda \Delta^\lambda_i \rangle_s = \langle \Big(\xi_i^\nu \frac{\partial g^{(i, \lambda)}_\nu(x_i, g_\lambda)}{\partial x_i}\biggr \rvert_{0,0}\Big)\cdot \Big(\xi_i^\lambda  \frac{\partial g^{(i, \nu)}_\lambda(x_i, g_\nu)}{\partial x_i}\biggr \rvert_{0,0}\Big)\rangle_s + O(1/N) = \langle \xi_i^\nu \Delta^\nu_i \rangle_s \langle \xi_i^\lambda \Delta^\lambda_i \rangle_s + O(1/N)$}. The values $\langle \xi_i^\nu \Delta^\nu_i \rangle_s$ are averaged over the randomly drawn strategies, and hence independent of $\nu$ and $i$. The value of $R_x$ is therefore also independent of $i$. The interpretation of the three terms is as follows:
$z_{i \setminus \mu}$ is the cavity field in absence of $g_{\mu}$,  $c_{\mu i}$ is an effective interaction,
and $R_x x_i$ is the reaction of the market according to (A.2) to agent $i$ (i.e., the market impact of agent $i$).
We can solve Eq.~\ref{eq: cavity_eq_x_app} for $x_i$ and obtain:
\begin{align} \label{eq: x_i_2}
    x_i^{(\mu)}(g_{\mu}) &= \hat{x}(z_{i \setminus \mu} + c_{\mu i}g_{\mu} + \epsilon_i) \,, \\
    x_i^\ast = x_i^{(\mu)}(g^\ast_{\mu}) &= \hat{x}(z_i + \epsilon_i)\,, \label{eq: xast_eps}
\end{align}
with $\hat{x}(z_x)$ as in the main text (Eq.~\ref{eq: xhat}) and the cavity field:
\begin{align}
    z_i \equiv \frac{1}{\sqrt{N}} \sum_{\nu}  \xi_i^\nu g_{\nu \setminus i} = z_{i \setminus \mu} + c_{\mu i}g^\ast_{\mu} \,. \label{eq: z_i}
\end{align}

\subsubsection{Equation for the market to determine $g_\mu^*$}

To find $g^\ast_{\mu}$, we insert assumption 2 (Eq.~\ref{eq: assumption_2}) into Eq.~\ref{eq: remain_eq_A_mu} to obtain
\begin{align} \label{eq: A_eq}
    A_{\mu} &= (z_{\mu \setminus i} + c_{i\mu} x_i + \epsilon_\mu) + R_g g_{\mu} + O(1/N)\,,
\end{align}
where we defined
\begin{align}
    z_{\mu \setminus i} &\equiv \frac{1}{\sqrt{N}}\omega_i^\mu + \frac{1}{\sqrt{N}}\sum_{j \neq i} (\omega_j^\mu + \xi_j^\mu x_{j \setminus i,\mu})  \,, \label{eq: zmu/i}\\
    c_{i\mu} &\equiv \frac{1}{\sqrt{N}} \xi_i^\mu + \frac{1}{\sqrt{N}}\sum_{j \neq i} \xi_j^\mu \Delta^j_i \label{eq: c_imu}\\
    R_g &\equiv \frac{1}{\sqrt{N}}\sum_{j} \xi^\mu_j \Delta^{j}_{\mu} + O(1/N)\\
     &=  \frac{1}{\sqrt{N}}\sum_{j} \langle \xi^\mu_j \Delta^{j}_{\mu} \rangle_s  + O(1/\sqrt{N}) \,,
\end{align}
where in the last line we used the central limit theorem\footnote{The terms are up to $O(1/N)$ uncorrelated, since for agents $j$ and $k$, $\langle \xi^\mu_j \Delta^{j}_{\mu} \cdot \xi^\mu_k \Delta^{k}_{\mu}\rangle_s = \langle \Big(\xi^\mu_j \frac{\partial x_{j}^{(k,\mu)}(x_k, g_\mu)}{\partial g_\mu}\biggr \rvert_{0,0} \Big) \cdot \Big(\xi^\mu_k \frac{\partial x_{k}^{(j,\mu)}(x_j, g_\mu)}{\partial g_\mu}\biggr \rvert_{0,0}\Big)\rangle_s + O(1/N) = \langle \xi^\mu_j \Delta^{j}_{\mu} \rangle_s \langle \xi^\mu_k \Delta^{k}_{\mu}\rangle_s + O(1/N)$.} to show that the value of $R_g$ is independent of $\mu$. Note that $c_{i\mu} \neq c_{\mu i}$. Solving for $g_{\mu}$ then gives the equation:
\begin{align} \label{eq: A_mu_2}
    g^{(i)}_{\mu}(x_i) &=  \hat{g}(z_{\mu \setminus i} + c_{i \mu}x_i + \epsilon_\mu)  \\
    g_{\mu}^\ast = g^{(i)}_{\mu}(x^\ast_i) &= \hat{g}(z_\mu + \epsilon_\mu) \,,  
\end{align}
with $\hat{g}(z_g)$ as in the main text (Eq.~\ref{eq: ghat}) and the cavity field:
\begin{align}
    z_{\mu} \equiv \frac{1}{\sqrt{N}}\sum_j (\omega_j^\mu + \xi_j^\mu x_{j \setminus \mu}) = z_{\mu \setminus i} + c_{i \mu}x^\ast_i \label{eq: z_mu}
\end{align}

\subsubsection{Distribution of the cavity fields}

Note that the reaction terms take fixed values as a function of the input parameters, while the cavity fields are random variables, for which we determine next their Gaussian distributions. Now that the correlations between the agents and the market have been accounted for by the reaction terms, the cavity fields $z_i$ and $z_\mu$ are large sums of independent random variables, and thus we can apply the central limit theorem to determine they are normally distributed. We have:
\begin{align}
    \langle z_i \rangle_s = \frac{1}{N} \sum_{\mu} \langle \xi_i^\mu g_{\mu \setminus i}\rangle_s =  \frac{1}{N} \sum_{\mu} \langle \xi_i^\mu \rangle_s  \langle g_{\mu \setminus i} \rangle_s = 0
\end{align}
and
\begin{align}
    \langle z_i^2 \rangle_s &= \frac{1}{N} \sum_{\mu}  \langle (\xi_i^\mu)^2  g_{\mu \setminus i}^2 \rangle_s \\
    &= \frac{1}{N} \sum_{\mu}  \langle (\xi_i^\mu)^2 \rangle_s \langle g_{\mu \setminus i}^2 \rangle_s\\
    &= \frac{1}{2N} \sum_{\mu} \langle g_{\mu \setminus i}^2 \rangle_s \\
    &= \frac{\alpha}{2}q_g + O(1/\sqrt{N}) \equiv \langle z_x^2 \rangle_s\,,\label{eq: neglect}
\end{align}
\noindent with $q_g \equiv \frac{1}{P} \sum_{\mu} (g^\ast_{\mu})^2$, and where we defined $\langle z_x^2 \rangle_s$ because $\langle z_i^2 \rangle_s$ is independent of $i$.
The difference between $g^\ast_{\mu}$ and $g_{\mu \setminus i}$ is of order $\frac{1}{\sqrt{N}}$ and hence we have neglected it.
It is, however, crucial that we did this only after decoupling $\langle (\xi_i^\mu)^2 g_{\mu \setminus i}^2 \rangle_s = \langle (\xi_i^\mu)^2 \rangle_s \langle g_{\mu \setminus i}^2 \rangle_s $:
Since $g_{\mu \setminus i} = g_{\mu}^{(i)}(0)$ is evaluated at a fixed value $x_i = 0$, it is uncorrelated to $\xi_i^\mu$, which is not true for $g^\ast_{\mu} = g_{\mu}^{(i)}(x_i^\ast)$ (since $x_i^\ast$ is correlated to $\xi_i^\mu)$.
As we discussed in Sec.~\ref{sec: CLT}, without circumventing these correlations, we would not be allowed to use the central limit theorem.
Now that we have been able to use the central limit theorem to find the distribution of $z_i$,
the distinction between $g_{\mu \setminus i}$ and $g_{\mu}^\ast$ has served its purpose (it has introduced the reaction term $R_x$), and we are allowed to neglect the difference for the distribution $P(z_x)$.

Similarly, the cavity field $z_{\mu}$ is  a Gaussian random variable with mean
\begin{align}
    \langle z_{\mu} \rangle_s = \frac{1}{\sqrt{N}}\sum_j \langle \omega_j^\mu + \xi_j^\mu x_{j \setminus \mu}\rangle_s &= \frac{1}{\sqrt{N}}\sum_j \big(\langle \omega_j^\mu\rangle_s + \langle \xi_j^\mu \rangle_s \langle x_{j \setminus \mu}\rangle_s  \big) \\
    &= b = \langle z_g \rangle_s \,,
\end{align}
where we used $\langle \omega_j^\mu \rangle_s = b/\sqrt{N}
$, and variance
\begin{align}
    \langle (z_{\mu} - b)^2 \rangle_s &= \frac{1}{N} \sum_j \big(\langle(\omega_j^\mu)^2\rangle_s + \langle(\xi_j^\mu)^2 x_{j \setminus \mu}^2\rangle_s \big) \\
    &= \frac{1}{N} \sum_j \big(\langle(\omega_j^\mu)^2\rangle_s + \langle(\xi_j^\mu)^2\rangle_s \langle x_{j \setminus \mu}^2\rangle_s \big) \\
    &= \frac{1}{2} + \frac{1}{2}q_x + O(1/\sqrt{N}) = \langle (z_g - b)^2 \rangle_s \,,
\end{align}
where $q_x$ is as in Eq.~\ref{eq: q_x} and we defined $\langle z_g \rangle_s$ and $\langle (z_g - b)^2 \rangle_s$ because $\langle z_\mu \rangle_s$ and $\langle (z_\mu - b)^2 \rangle_s$ are independent of $\mu$. In the end we could again neglect the difference between $x_{j \setminus \mu}$ and $x^\ast_j$ as being of order $O(1\sqrt{N})$; once again we could only do this after decoupling $\langle (\xi_j^\mu)^2 x_{j \setminus \mu}^2 \rangle_s  = \langle (\xi_j^\mu)^2 \rangle_s \langle x_{j \setminus \mu}^2 \rangle_s$.

\subsubsection{The reaction terms}

From Eqs.~\ref{eq: x_i_2} and \ref{eq: A_mu_2} we get:
\begin{align}
    \Delta^i_{\mu} &= c_{\mu i} \hat{x}'(z_{i \setminus \mu} + \epsilon_i) = c_{\mu i} \hat{x}'(z_{i}) + O(1/N) \,, \label{eq: Delta_i_mu}\\
    \Delta^\mu_i &= c_{i \mu} \hat{g}'(z_{\mu \setminus i} + \epsilon_\mu) = c_{i \mu} \hat{g}'(z_\mu) + O(1/N) \label{eq: Delta_mu_i}\,,
\end{align}
where the apostrophe denotes the derivative. Since an increase in $z_\mu$ corresponds to an increase in $A_\mu$ (Eq.~\ref{eq: cavity_eq_A}), which in turn increases $g_\mu$, we have $\hat{g}' \geq 0 $. On the other hand, an increase in $z_i$ corresponds to a decrease in $\overline{U^t_i}$ (Eqs.~\ref{eq: Ut/t}, \ref{eq: stat_eq_3} and \ref{eq: cavity_eq_x}), so $\hat{x}' \leq 0$. We  use the expressions for $\Delta^i_{\mu}$ and $\Delta^\mu_i$ to find the reactions $R_x$ and $R_g$ from Eqs.~\ref{eq: R_x} and \ref{eq: R_g}.
To find $R_g$, we should notice that all terms in $c_{\mu i}$ involving $\Delta^\nu_{\mu}$ are of order $1/\sqrt{N}$
and uncorrelated with $\xi_j^\mu$.
Therefore, they do not contribute to $R_{g}$, and the expression for $\Delta^j_{\mu}$ (Eq.~\ref{eq: Delta_i_mu}) gives:
\begin{align}
    R_g &= \frac{1}{\sqrt{N}}\sum_j \langle \xi_j^\mu \Delta^j_{\mu} \rangle_s + O(1/\sqrt{N})\\
    &= \frac{1}{N}\sum_j \langle  (\xi_j^\mu)^2 \rangle_s \langle \hat{x}'(z_\mu)\rangle_s + O(1/\sqrt{N})
    = \frac{1}{2}  \int \text{d}z_x \, \hat{x}'(z_x) P(z_x) + O(1/\sqrt{N}) \leq 0 \,.  \label{eq: no_corr}
\end{align}
We can also use the expression for $\Delta^\mu_i$ to calculate $R_x$.
Finding $c_{i \mu}$ from Eq.~\ref{eq: c_imu}, the terms involving $\Delta^j_i$ are of order $1/\sqrt{N}$ and uncorrelated to $\xi_i^\mu$ and $\xi_j^\mu$,
so that we can neglect them:
\begin{align}
    R_x &= - \frac{1}{\sqrt{N}}\sum_{\mu} \langle \xi_i^\mu \Delta^\mu_i \rangle_s \label{eq: R_eq_2} \\
    &= - \frac{1}{N} \sum_{\mu} \langle(\xi_i^\mu)^2\rangle_s \langle \hat{g}'(z_\mu) \rangle_s = -\alpha/2 \int \text{d}z_g \, \hat{g}'(z_g) P(z_g) \leq 0 \,,\label{eq: R_eq_1}
\end{align}
\subsection{Self-consistent verification of the cavity assumptions}

We obtained $x_i^{(\mu)}(g_\mu)$ and $g_\mu^{(i)}(x_i)$ (Eqs.~\ref{eq: x_i_2} and \ref{eq: A_mu_2}), and found that they are of the form of Eq.~\ref{eq: x_imu}. To fully verify the cavity assumptions we must also find a self-consistent expression for the more general terms $x_j^{(i,\mu)}(x_i, g_\mu)$ and $g_\nu^{(i, \mu)}(x_i, g_\mu)$ in a way that we can explicitly read off their order of magnitude. To this end we should notice that, inserting the assumption of Eq.~\ref{eq: cavity_abbrv_1} into the equation for agent $j$ (from Eq.~\ref{eq: stat_eq_eps_i}), we get that the equation for $x_j^{(i,\mu)}(x_i,g_\mu)$ becomes:
\begin{alignat}{1}
    & -\frac{1}{\sqrt{N}}\sum_{\nu} \xi_j^\nu g^{(i,\mu)}_{\nu}(x_i, g_\mu) - \epsilon_j \\
    &= -\frac{1}{\sqrt{N}}\sum_{\nu} \xi_j^\nu g_{\nu \setminus i, \mu} - \epsilon^{(i,\mu)}_j(x_i, g_\mu) =
    \begin{cases}
        0  &\text{if} \quad -1 < x_j^{(i,\mu)}(x_i,g_\mu) < 1  \\
        > 0 & \text{if} \quad x_j^{(i,\mu)}(x_i,g_\mu) = 1
        \\
        < 0 & \text{if} \quad x_j^{(i,\mu)}(x_i,g_\mu) = -1
    \end{cases} \,,
\end{alignat}
where
\begin{alignat}{1}
    \epsilon_j^{(i,\mu)}(x_i, g_\mu) &= (\epsilon_j  + c_{ij} x_i + c_{\mu j} g_\mu) \,,\label{eq: eps_adapted}\\
    c_{ij} &= \frac{1}{\sqrt{N}} \sum_{\nu} \xi_j^\nu \Delta^\nu_i \,,
\end{alignat}
and with $c_{\mu j}$ as in Eq.~\ref{eq: c_mui}. The only dependence of $x_j^{(i,\mu)}(x_i,g_\mu)$ on $x_i$ and $g_\mu$ is due to the adapted perturbation $\epsilon_j \rightarrow \epsilon^{(i,\mu)}_j(x_i,g_\mu)$. With this adapted perturbation, the solution of the $N' = N-1$ and $P' = P - 1$ equations for agents $k \neq i$ and signals $\nu \neq \mu$ thus gives $x_j^{(i,\mu)}(x_i,g_\mu)$.
Since Eq.~\ref{eq: xast_eps} also holds for any agent in the system of $N' + P'$ equations, including agent $j$, we have:
\begin{align}
    x^{(i, \mu)}_j(x_i, g_\mu) = x_{j \setminus i, \mu} + \hat{x}'(z_j) \cdot (\epsilon^{(i,\mu)}_j  (x_i,g_\mu) - \epsilon_j) + O(1/N),
\end{align}
which implies:
\begin{alignat}{1}
    x^{(i, \mu)}_j(x_i, g_\mu)
    &= x_{j \setminus i, \mu} +  \hat{x}'(z_{j}) \cdot(c_{ij}x_i + c_{\mu j}g_\mu)\,. \label{eq: x(xi, gmu)}
\end{alignat}
The same argument for $g^{(i,\mu)}_\nu(x_i, g_\mu)$ gives:
\begin{alignat}{1}
    g^{(i, \mu)}_\nu(x_i, g_\mu)
    &= g_{\nu \setminus i, \mu} + \hat{g}'(z_{\nu}) \cdot (c_{i \nu}x_i + c_{\mu \nu}g_\mu) \label{eq: g(xi, gmu)}\,,
\end{alignat}
with $c_{i \nu}$ as in Eq.~\ref{eq: c_imu} and $c_{\mu \nu} = \frac{1}{\sqrt{N}} \sum_i \xi_i^\nu \Delta^i_{\mu}$.
The cavity assumptions thus self-consistently confirm the expressions of Eqs.~\ref{eq: cavity_abbrv_1}-\ref{eq: cavity_abbrv_2} with:
\begin{equation} \label{eq: Deltas}
    \begin{aligned}[c]
        \Delta^\nu_i &= c_{i \nu} \cdot \hat{g}'(z_\nu) \,, \\
        \Delta^j_\mu &= c_{\mu j} \cdot\hat{x}'(z_j) \,,
    \end{aligned}
    \qquad\qquad
    \begin{aligned}[c]
        \Delta^\nu_\mu &= c_{\mu \nu} \cdot \hat{g}'(z_\nu) \,,\\
        \Delta^j_i  &= c_{ij} \cdot \hat{x}'(z_j) \,,
    \end{aligned}
\end{equation}
with the effective interaction coefficients (all of order $1/\sqrt{N}$):
\begin{equation} \label{eq: coeffs}
    \begin{aligned}[c]
        c_{i\nu} &= \frac{1}{\sqrt{N}} \xi_i^\nu + \frac{1}{\sqrt{N}}\sum_{j \neq i} \xi_j^\nu \Delta^j_i \,, \\
        c_{\mu j} &= \frac{1}{\sqrt{N}}\xi_j^\mu + \frac{1}{\sqrt{N}}\sum_{\nu \neq \mu} \xi_j^\nu \Delta_{\mu}^\nu \,,
    \end{aligned}
    \qquad\qquad
    \begin{aligned}[c]
        c_{\mu \nu} &= \frac{1}{\sqrt{N}} \sum_i \xi_i^\nu \Delta^i_{\mu} \,,  \\
        c_{ij}  &= \frac{1}{\sqrt{N}} \sum_{\mu} \xi_j^\mu \Delta^\mu_i \,.
    \end{aligned}
\end{equation}

\section{Notations and variables} \label{app: variables}

{\renewcommand{\arraystretch}{1.2}
\begin{longtable}{l l l}
    \hline
    Variable & Meaning & Value \\
    \hline
    {\small{Input parameters}} &  & \\
    $N$ & Number of agents & $\gg 1$ \\
    $P$  & Amount of possible values of the information & $ O(N)$ \\
    $S$ & Amount of strategies available to each agent & $\geq 2$ \\
    $P(\eta)$ & Distribution of external noise & Zero mean \\
    $g$  & Price function & $g' > 0$ \\
    $b$ & Biasing parameter for $a = \pm 1$ & s.t. $\overline{g_t} = 0$ \\
    &  & \\
    {\small{Averages}}  &  & \\
    $\overline{\mathbbm{O}_t}$& Average over time & $ \lim_{t \rightarrow \infty} \frac{1}{t}\sum_{t' \leq t} \mathbbm{O}_{t'} $\\
    $\langle \mathbbm{O}_{\{s\}} \rangle_s $& Average over randomly drawn strategies & $ \sum_{\{s\}} \mathbbm{O}_{\{s\}} p(\{s\}) $\\
    $\langle \mathbbm{O}_\mu(\eta) \rangle_{\mu, \eta} $  & Average over $\mu$ and $\eta$  &  $\frac{1}{P} \sum_{\mu = 1}^P \int \mathrm{d} \eta P(\eta)\mathbbm{O}_{\mu}(\eta) $\\
    $\big\langle\mathbbm{O}(\delta, \eta)\big\rangle_{\delta,\eta}$ & Average over noise terms $\eta$ and $\delta$ & $\int \mathrm{d}\delta \mathrm{d}\eta \mathbbm{O}(\delta, \eta) P(\delta)P(\eta)$ \\
    &  &  \\
    {\small{Microscopic variables}} \\
    $a_i^t$  & Arbitrage action of agent $i$ at time $t$ & $ \pm 1$ \\
    $s_i(\mu)$ & Strategy $s$ of agent $i$ given a signal $\mu$ & $\pm 1$\\
     $U_s^t$ & score of strategy $s$ at time $t$ &  $\sum_{t' \leq t} -s(\mu^{t'})g(A^t + \eta^t)$ \\
     $s_i^\uparrow$, $s_i^\downarrow$  & The two strategies available to agent $i$ & \\
     $\omega_i^\mu$ &  & $\frac{s_i^\uparrow(\mu) + s_i^\downarrow(\mu)}{2}$ \\
     $\xi_i^\mu$  &  & $\frac{s_i^\uparrow(\mu) - s_i^\downarrow(\mu)}{2}$\\
     $U_i^t$  & Difference in the scores between agents $i$'s strategies & $U_{s_i^\uparrow}^t - U_{s_i^\downarrow}^t$\\
     $x_i^t$  & Decision between strategies of agent $i$ at time $t$ & $\text{sign}(U_i^t)$\\
     $x_i$ & Average decision of agent $i$ & $\overline{x_i^t}$ \\
     $A_\mu^t$  & $A^t$ if $\mu^t = \mu$ & $ \frac{1}{\sqrt{N}}\sum_i (\omega_i^{\mu}  + \xi_i^{\mu} x^t_i)$\\
     $A_\mu$ & Time-average of $A_\mu^t$ & $\overline{A_\mu^t}$ \\
     $g_\mu$ & Time-average of $g(A_\mu^t + \eta^t)$ & $\overline{g(A_\mu^t + \eta^t)}$\\
     &  &\\
    {\small{Macroscopic variables}} \\
    $z_i$ & Cavity field for $x_i$ & Eq.~\ref{eq: z_i}\\
    $z_\mu$& Cavity field for $g_\mu$ & Eq.~\ref{eq: z_mu}\\
    $R_x$ & Reaction term for $\{x_i\}$ & Eq.~\ref{eq: R_x}\\
    $R_g$ & Reaction term for $\{g_\mu\}$ & Eq.~\ref{eq: R_g}\\
    $q_x$  & Measure of non-uniformity of $\{x_i\}$ & $\frac{1}{N}\sum_i x_i^2$\\
    $q_A$  & Measure of non-uniformity of $\{A_\mu\}$  & $\frac{1}{P} \sum_\mu (A_\mu - b)^2$\\
    $q_g$ & Measure of non-uniformity of $\{g_\mu\}$ & $\frac{1}{P} \sum_\mu g_\mu^2$\\
    $1 - \phi$ & Fraction of non-frozen agents & $\frac{1}{N}\sum_i \mathbbm{1}(-1 < x_i < 1)$ \\
    $\sigma$ & Volatility & $\sqrt{\overline{(A^t + \eta^t - b)^2}}$
\end{longtable}}
\end{document}